\documentclass[aps,a4paper,floatfix,pre,showpacs,preprint,merge,elide]{revtex4}
\usepackage{graphicx}
\usepackage[dvipsnames]{pstricks}
\usepackage{amsmath,amsfonts,amssymb}
\def\bg#1{\hbox{\bf#1}}

\newcommand{\be}{\begin{eqnarray}}
\newcommand{\ee}{\end{eqnarray}}
\newcommand{\bC}{\begin{center}}
\newcommand{\eC}{\end{center}}
\newcommand{\befi}{\begin{figure}}
\newcommand{\enfi}{\end{figure}}
\newcommand{\bi}{\bibitem}
\newcommand{\ci}{\cite}
\newcommand{\la}{\label}

\newcommand{\benl}{\begin{eqnarray*}}
\newcommand{\eenl}{\end{eqnarray*}}
\newcommand{\REM}[1]{}

\DeclareMathOperator{\Tr}{tr}
\makeatletter
\newcommand\mysection{\@startsection{section}{1}{\z@}%
                                   {-3.5ex \@plus -1ex \@minus -.2ex}%
                                   {2.3ex \@plus.2ex}%
                                   {\normalfont\large\sffamily\centering}}
\newcommand\mysubsection{\@startsection{subsection}{1}{\z@}%
                                   {-3.5ex \@plus -1ex \@minus -.2ex}%
                                   {2.3ex \@plus.2ex}%
                                   {\normalfont\sffamily\centering}}
\makeatother
\begin{document}
\title{{\sf Computer simulation study} 
\\
{\sf of a mesogenic lattice model}
\\
{\sf based on long-range dispersion interactions}
}
\author{Silvano Romano}
\affiliation{Physics dept., the University,
via A. Bassi 6, 27100 Pavia, Italy \\
Silvano.Romano@pv.infn.it}

\parindent 0cm
\begin{abstract}
In contrast  to thermotropic biaxial nematic phases,
for which some long sought for experimental realizations have been obtained, 
no experimental realizations are yet known for
their tetrahedratic and cubatic counterparts,
involving orientational orders of ranks 3 and 4, respectively,  also studied
theoretically over the last few decades. 
In previous studies,
cubatic order has been found for hard-core or continuous models
consisting of particles possessing  cubic or nearly-cubic tetragonal or orthorhombic  symmetries;
in a few cases, hard-core models involving uniaxial ($D_{\infty h}-$symmetric) 
particles have been claimed to produce cubatic order as well.
Here we address by Monte Carlo simulation a lattice model consisting of uniaxial
particles coupled by long-range dispersion interactions
of the London-De Boer-Heller type;
the model was found to produce no second-rank nematic but only fourth-rank cubatic order,
in contrast to the nematic behavior long known for its counterpart with interactions
truncated at nearest-neighbor separation.
\end{abstract}
\date{\today}
\maketitle


\parindent 0cm
\section{Introduction} \label{intro}
Thermotropic biaxial nematic liquid crystals 
have been called "the holy grail" of liquid crystal research,
both for fundamental reasons and in connection with their
possible technological applications in displays \ci{rGRLnature};
 a coherent pedagogical account 
of background and present state of the art on the subject has been  published in 2015 
\ci{book00}, and we refer to it
for a more detailed discussion and more extensive bibliography of the different aspects of 
this fascinating and challenging subject.

Here we briefly recall that nematic phases are usually
apolar and uniaxial ($D_{\infty h}$-symmetric), although the constituent 
molecules possess lower symmetries;  biaxiality was 
discovered in a smectic C phase in 1970
\ci{prl02400359}, and, in the same year, 
Freiser \ci{prl02401041} addressed
 the possibility of biaxial mesogenic molecules 
producing  a biaxial  nematic phase;
the following years saw extensive theoretical and simulational investigations
to elucidate the properties of the hypothetical biaxial phase 
(especially in theoretical work, molecular $D_{2h}$ symmetry
has mostly been studied, but  lower symmetries have been addressed as well
in recent years \ci{book15}).
There also followed  several attempts to produce experimental realizations,
which remained unsuccessful until the end of the past century;
better evidence was obtained over the last twelve years  \ci{book14,book15};
in these cases, second-rank orientational order is involved.

%
%
On the other hand, the theoretical possibility of positional disorder accompanied by
orientational order of other point-group symmetries, involving  
tensors of rank $L$ different from 2, has been investigated theoretically 
 for some 30 years to date
\ci{npb26500647,pre05200702,pre05202692,pre074041701}, 
especially for tetrahedratic ($L=3$) and cubatic orders ($L=4$);
a detailed symmetry classification of ``unconventional'' nematic phases,
i.e. associated with the onset of either one tensor of rank
different from 2 or of several combined tensors,  
has   been carried out by Mettout \ci{pre074041701};
a general classification of point-group symmetric orientational
ordering tensors has recently been published \ci{pre094022701}.
This line of investigations has partly overlapped with  the study
of  packings of hard nonspherical particles, including a number of   polyhedra
(see, e.g. Refs. \ci{rev001,rev002,rev003,rev004,rev005,rev006,rev007}):
 in addition to the extensive search for the densest
packing obtainable for  a specific  particle geometry, resulting
phase diagrams (including possible mesophases) have  been  studied.

The possibility of tetrahedratic orientational order, involving
a third--rank order parameter was proposed and studied 
by L. G. Fel \ci{pre05200702,pre05202692}; its transitional behavior was 
later studied by Lubensky and Radzihovsky \ci{rLR01,rLR02},
and the macroscopic consequences of tetrahedratic order were
 discussed in detail by Brand, Cladis and Pleiner \ci{rBCP01,rBCP02,rBCP03}.
Continuous interaction lattice models 
involving third-rank interactions alone \ci{rtetsim01} or combined
with second-rank terms \ci{rtetsim02,rtetsim03} and 
producing tetrahedratic order have also been studied by simulation.
Starting in the mid-1990's, bent-core (banana-shaped) mesogens
were synthesized  \ci{rbanrev}, and found to produce  
mostly  smectic, and   sometimes  nematic \ci{rbannem}
phases; 
to the best of our knowledge, no experimental realizations of a purely
tetrahedratic phase are known at the time being,
no third--rank order parameter has been measured to date,  
yet the theoretical analyses in Refs. \ci{rLR01,rLR02,rBCP01,rBCP02,rBCP03}
show that interactions of tetrahedral symmetry 
(or, in more general terms, a description allowing
for first, second and third--rank ordering  tensors)
are needed for the proper comprehension 
of macroscopic properties of mesophases resulting from
 bent--core molecules.
Some experimental evidence suggesting tetrahedratic behavior in an 
achiral bent-core liquid crystal has been reported in Ref. \ci{rtet03}.

\REM{ 
the abstract of Ref.  \ci{rtet03} reads
\begin{quote}
An experimental study of the heat capacity, mass density, magnetic-field-induced optical birefringence, linewidth and intensity of scattered light, and the viscosities associated with nematic order parameter fluctuations and fluid flow has been performed on an achiral bent-core liquid crystal above its clearing point temperature. The measurements reveal a transition between two optically isotropic phases that is consistent with recent theoretical predictions of a ``tetrahedratic'' form of orientational order.
\end{quote}
} 

%
%
Over the last decades, the possible existence of a 
cubatic mesophase,
 possessing cubic orientational order (i.e. along
three equivalent mutually orthogonal axes) but no translational one, 
has been extensively investigated by approximate analytical theories and by simulation
\ci{rcu00,rcu01,rcu02,rcu03,rcu04,rcu04add,rcu05,rcu06,rcu06add,pre074011704,
rcu07,rcu08,rcu09,rcu10},  and explicitly predicted in some cases;
hard-core models
possessing cubic or nearly-cubic tetragonal or orthorhombic  symmetries
have  been studied rather extensively:
for example Onsager crosses were studied in Ref. \ci{rcu04}, and tetrapods
were investigated in Ref. \ci{rcu04add};
arrays of hard spheres with tetragonal or cubic symmetries  have
been studied in Refs. \ci{rcu05,rcu06,rcu06add};
continuous interaction models possessing cubic symmetry have been studied
as well \ci{pre074011704}.
Hard-core models involving uniaxial particles have also been investigated
\ci{rcu00,rcu01,rcu02,rcu03,rcu07,rcu08,rcu09}: no cubatic order was found for
hard cylinders \ci{rcu03}, whereas 
symmetrically cut hard spheres \ci{rcu00,rcu07,rcu08,rcu09} of appropriate 
length-to-width appear to produce a metastable cubatic phase.

As hinted above, no  experimental realizations of a thermotropic cubatic phase are
known at the time being; on the other hand, liquid crystal phase transitions
in suspension of mineral colloids \ci{rLV} have been investigated for some
ninety years to date,  and evidence of cubatic order has been proposed  in Ref. 
\ci{rQKR}, where uniform sterically stabilized hexagonal platelets
of nickel(II) hydroxide had been dispersed in $D_2O$.
A few years later, Molecular Dynamics simulations \ci{rcu10}
addressed colloidal platelets with a square cross-section, and consisting
of fused spherical interaction centers; a stable cubatic phase was reported.

It also seems appropriate to recall that, 
starting with the seminal Lebwohl-Lasher simulation papers in the early 1970's
\ci{rLL01,rLL02}, 
mesophases possessing no 
positional order, such as the nematic one(s), have  often 
been studied by means of lattice models
involving continuous interaction potentials \ci{bpz-05,rBL}; 
this approach also 
yields a convenient  contact with Molecular Field
(MF) treatments of the Maier-Saupe (MS) type \ci{rms,rmfb,bqz-00}. 
It has been pointed out \ci{rBL}
that  usage of a 
lattice model reduces the number of parameters to be controlled, thus 
producing   important  savings in computer time, and, moreover, 
it excludes a  number a competing phases (e.g. smectic ones) 
from the start;  similar simplifications 
as for the possible phases  are  used  
in  other theoretical treatments as well.

The present communication reports a  Monte Carlo (MC) simulation
study, addressing and revisiting a  lattice model involving 
uniaxial particles, 
coupled via  a long-range dispersion potential, and producing cubatic
but no nematic order.

The rest of this paper is organized as follows: 
the interaction potential and its ground state are recalled 
in Section \ref{ground};
simulation aspects are  briefly discussed in Section 
\ref{comptaspect}; simulation results are presented in Section \ref{results}, 
and the paper is concluded 
in Section \ref{concl}, where results are summarized.

\newpage
\parindent 0cm
\section{Interaction Model  and  Ground State} \label{ground}
 As for symbols and definitions,
we are considering here 
 $3-$component unit vector (classical spins),
associated with the nodes of a $3-$dimensional simple-cubic lattice $\mathbb{Z}^3$;
let $\mathbf{x}_j$ denote the coordinate vectors of lattice sites,
let $\mathbf{w}_j$ denote the unit vectors, and let $w_{j,\iota}$
denote their Cartesian components with respect to an orthonormal basis
$\mathcal{E}=\{\mathbf{e}_{\iota},~\iota=1,2,3\}$ defined by lattice axes;
 the unit vectors $\mathbf{w}_j$ can also
be parameterized by usual polar angles $\left( \theta_j,~\phi_j \right)$.

The quantum theory
of intermolecular forces
\ci{rimfAA,rimfBB} predicts for
the dipolar contribution to the dispersion energy
between two identical, neutral and centrosymmetric
linear molecules the general form  
\be
\Delta_{jk}^0=
\frac{1}{r^6} \left[
g_0+g_1 (a_j^2+a_k^2) + g_2  a_j a_k b_{jk}
+g_3 b_{jk}^2 + g_4  (a_j a_k)^2 \right],  
\la{eq003}
\ee
where 
\be
\mathbf{r} = \mathbf{r}_{jk} =\mathbf{x}_j-\mathbf{x}_k,~r=|\mathbf{r}|,~\hat{\mathbf{r}}=\mathbf{r}/r,~
\la{eq004}
\\
a_j = \mathbf{w}_j \cdot \hat{\mathbf{r}},~
a_k = \mathbf{w}_k \cdot \hat{\mathbf{r}},~b_{jk}=
\mathbf{w}_j \cdot \mathbf{w}_k,   
\la{eq005} 
\ee
and the $g$ coefficients can be calculated based on the unperturbed wave functions.
Under additional simplifying approximations, Eq. (\ref{eq003})
leads   to the expression proposed 
by London, de Boer and Heller ($LBH$) in the $1930's$
 \ci{rdispa,rdispb,rdispc,rdispd}, i.e. 
\be
\Delta_{jk}=\frac{\epsilon}{ r^6} 
[(\gamma^2 - \gamma) S_{jk}
- \frac{3}{2} \gamma^2 h_{jk} +\gamma^2-1],~ 
\epsilon=\frac{3}{4} \overline{E} \overline{\alpha}^2,
\la{eq006} 
\ee
where 
\be
h_{jk}=(3 a_j a_k -b_{jk})^2,~S_{jk}=P_2(a_j)+P_2(a_k),~
\overline{\alpha}=\frac{1}{3}(\alpha_{\parallel}+2 \alpha_{\perp}),~
\gamma=\frac{\alpha_{\parallel}-\alpha_{\perp}}{3 \overline{\alpha}}.
\la{eq009}
\ee
Here 
$P_2(\ldots)$ denote second Legendre polynomials of the relative
arguments, 
$\alpha_{\parallel},\alpha_{\perp}$ are 
the eigenvalues of the molecular polarizability tensor, 
$\gamma$ denotes its relative anisotropy,
and  $\overline{E}$ is a mean excitation energy;
formulae are also known for higher-order terms in the 
multipolar expansion  \ci{rimfAA,rimfBB,rdispd};
the extreme case $\gamma=-(1/2)$ corresponds to no polarizability
along the molecular symmetry axis, whereas  
in the other extreme
$\gamma=+1$ 
there is polarizability along the molecular  axis only.
In the following, let $\tilde{\Delta}_{jk}$ denote the restriction
of $\Delta_{jk}$ to nearest neighbors (n-n), i.e.
\be
\tilde{\Delta}_{jk}= \epsilon 
[(\gamma^2 - \gamma) S_{jk}
+\gamma^2
(- \frac{3}{2}  h_{jk} +1)],~ 
\la{eq006nn} 
\ee
where a  purely positional and $\gamma-$independent term appearing in 
Eq. (\ref{eq006}) has been dropped.

Eq. (\ref{eq003}) or (\ref{eq006}) 
have  been used  in the literature,
usually as one  component of  the pair potential between
comparatively simple linear molecules
(see, e.g., Refs. \ci{rdex00,rdex01,rdex03});
limitations and possible improvements of the $LBH$
interaction model have  been discussed in the
Literature as well (see, e.g., Refs. \ci{rimfBB,robj}).

Models based on Eq. (\ref{eq006nn}) have  
been investigated as possible mesogens by simulation, 
both on a 3-dimensional ($3-d$)
\ci{rdisp1,PSthesis,rdisp2,
rBates} and on  a $2-d$  lattice \ci{r2ddisp};
on a $3-d$ lattice,
the n-n model $\tilde{\Delta}_{jk}$  
 was found to produce a nematic-like ordering 
transition \ci{rdisp1,rBates}; on the other hand,
 inclusion of next-nearest neighbors had been 
found to produce  a staggered ground state structure with sub-lattice 
order but no net second-rank  orientational order \ci{PSthesis,
rdisp2} (the $D_3$-type configurations mentioned below).

Another related mesogenic potential model,
proposed by Nehring and Saupe ($NS$) \ci{rsi00},
has the form
\be
\Gamma_{jk}=
- \frac{\epsilon}{r^6} h_{jk};
\la{eq010}
\ee
it has  been used for approximate 
calculations of elastic properties \ci{rJeu,rBBel};
its restriction to n-n, defined by 
\be
\tilde{\Gamma}_{jk}=\epsilon (-\frac{3}{2} h_{jk} +1),
\la{eq010nn}
\ee
has later been studied by simulation in $3-d$ \ci{rHR,rselflgns} as well as
in $2-d$ \ci{r2dns}.
Comparison between the relevant equations (Eq. (\ref{eq006}) and 
(\ref{eq010})),
shows that $NS$ corresponds to the limiting case $\gamma=+1$
in the $LBH$ model;
actually,
on a saturated cubic lattice and under periodic boundary conditions,
the two  models $\Delta_{jk}$ and $\Gamma_{jk}$ 
become equivalent within purely positional terms.
More explicitly, consider n-n interactions, 
in a periodically repeated sample, where
each particle interacts with 6  nearest neighbors only, 
and the possible orientations of the intermolecular
vector $\hat{\mathbf{r}}$ are $\pm {\bg e}_{\iota},~\iota=1,2,3$;
let us also  recall that,
for any unit vector 
$\mathbf{w}_j$ and for any lattice site $\mathbf{x}_j$,
\be
\mathbf{w}_j \cdot \mathbf{w}_j = \sum_{\iota=1}^{3} 
\left(\mathbf{e}_{\iota} \cdot \mathbf{w}_j\right)^2,~\sum_{\iota=1}^{3} 
P_2\left(\mathbf{e}_{\iota} \cdot \mathbf{w}_j\right)=0;
\la{eqiddisc}
\ee
this  identity
entails that, upon summing over all interacting pairs, the terms 
in the pair potential containing $S_{jk}$
 cancel out identically \ci{rBates,rHR}.

Moreover, 
let $m=h^2+k^2+l^2> 0$, denote the sum of squares of three integers,  and consider the sums
\be
c(m) =\sum_{\mathbf{r} \in \mathbf{Z}^3 \setminus \left\{\mathbf{0}\right\},~\mathbf{r} \cdot \mathbf{r} = m}
P_2\left(\mathbf{w} \cdot \hat{\mathbf{r}} \right),~\hat{\mathbf{r}} = \frac{\mathbf{r}}{|\mathbf{r}|};
\la{eqtr01}
\ee
then 
\be
c(m)=0,
\la{eqtr02}
\ee
for all $m$ and for any unit vector $\mathbf{w}$;
this result does  not only hold 
for a simple-cubic lattice, but also
for its body-centered (BCC)  and face-centered (FCC) counterparts;
thus, for a periodically repeated cubic sample, 
and for any truncation radius,
the  $S_{jk}$ terms 
in Eq. (\ref{eq006})
cancel out identically when  summed  over all interacting pairs, so that
terms linear with respect to $\gamma$ drop out, and only some terms proportional
to $\gamma^2$ survive; in other words, in the above setting, the dispersion model
$\Delta_{jk}$ (Eq. (\ref{eq006}))
and  the Nehring-Saupe model $\Gamma_{jk}$ (Eq. (\ref{eq010}))
 become equivalent, within purely distance-dependent terms;
the product  $\epsilon \gamma^2$ in Eq. (\ref{eq006}) or the quantity
$\epsilon$ in Eq. (\ref{eq010}) can be used to set
energy and temperature scales (i.e $T^*=k_B T/\epsilon$, where
$k_B$ denotes the Boltzmann constant).

To summarize, the present simulations, carried out on periodically
repeated cubic samples, used the functional form
\be
\Gamma_{jk}=
+ \frac{\epsilon}{r^6} \left(-\frac{3}{2} h_{jk} + 1 \right),
\la{eq010mod}
\ee
mostly with  truncation condition  $\mathbf{r} \cdot \mathbf{r} \le 25$.

Notice that   Eq. (\ref{eqtr02}), hence the
equivalence between the two potential models, do not hold for a sample being 
finite in some direction, e.g a $2-d$ lattice, nor for a grand-canonical (lattice-gas) simulation,
where each lattice site hosts one spin at most,
and its occupation number fluctuates
\ci{rBates,rselflgns}.


\REM{ 
For the model defined by
Eqs. (\ref{eq006}) or (\ref{eq006nn}) \ci{rdisp1,rBates,rHR},
 the pair configuration 
where the two molecules are parallel to each other and to the
intermolecular vector
(i.e. $a_j=a_k=b_{jk}= \pm 1$) has an energy
$-\epsilon(2 \gamma + 3 \gamma^2)/r^6$; in contrast, when both molecules 
are parallel to each other and perpendicular to the intermolecular vector,
i.e. $a_j=a_k=0,~b_{jk}=\pm 1$,
the energy is $\epsilon[\gamma -(3/2)\gamma^2]/r^6$, and 
both configurations correspond to minima in the dispersion potential;
these results may be   significantly modified when several longer-ranged  pair interactions 
are  involved, producing frustration.
} 

\REM{ 


3) Some of the results are reported from ref.64 from the same author
(e.g. page 8). Perhaps this part could be streamlined.

} 

A few  spin configurations possessing periodicity 2 in each lattice  
direction and constructed as in Ref.  \ci{rdisp2}
were examined as possible  ground state candidates,
 and the results further 
checked by simulations carried out  at low temperatures.

Procedure 
and definition of $D_1,~D_2,~D_3$ configurations are 
 recalled in Appendix \ref{appA} for readers' convenience;
let us notice that recognizing  cubatic order  in $D_3$-type configurations led to the present study.
\newpage
\parindent 0 cm
\section{Computational aspects}\label{comptaspect}

Simulations were carried out on a periodically repeated cubic  
sample, consisting of $N=l^3$ particles, 
$l=10,12,16,20,24$;
calculations were run in cascade, in order of increasing
temperature; 
each cycle (or sweep) consisted of $N$
$MC$ steps, and
the finest temperature step used was
 $\Delta T^*=0.001$, 
in the transition region.
Equilibration runs took between $25\,000$ and $100\,000$ cycles,
and production runs took between $500\,000$ and $3 \,500\,000$ (at least $2\,500\,000$ cycles in the transition
region, including the additional simulations mentioned below);
macrostep averages for evaluating statistical errors were 
taken over $1\,000$ cycles.  Calculated thermodynamic
quantities include mean
 potential energy per site  $U^*$ 
and  configurational specific heat per particle $C^*$,
where the asterisks
mean scaling by $\epsilon$ and $k_B$, respectively.

As for structural characterization,  we analyzed one configuration every cycle,
by calculating both
second- and fourth-rank
 ordering tensors  $\mathcal{T}^{(L)},~L=2,4$ \ci{rADB,r23,r24,bpz-02,SST}
as well as corresponding rotation-invariant
order parameters $O_L$. In   other words, for $L=2$,
\be
\mathcal{T}^{(2)}_{\iota \kappa}= Q_{\iota \kappa}= 
\frac{1}{2} (3 F_{\iota \kappa} - \delta_{\iota \kappa}),
\\
F_{\iota \kappa} = 
\frac{1}{N} \sum_{j=1}^N \left( w_{j,\iota} w_{j,\kappa}\right);
\ee
the calculated $\mathsf{Q}$ tensor can be diagonalized;  let $\left\{ q_k,~k=1,2,3\right\}$
denote its real eigenvalues, let $q^{\prime}$
denote the eigenvalue with maximum magnitude, and let $q_{max}$ denote
the maximum. eigenvalue; moreover 
\be
\mathsf{Q}: \mathsf{Q} = \Tr(\mathsf{ Q} \cdot \mathsf{Q}) = \sum_{k=1}^3 q_k^2,
\ee
where $:$ denotes the contracted product.
The fourth-rank counterpart is defined by
\be
\mathcal{T}^{(4)}_{\iota \kappa \lambda \mu}=
B_{\iota \kappa \lambda \mu} =
\nonumber
\\
  \frac{1}{8}
[35 G_{\iota \kappa \lambda \mu}
- 5 ( \delta_{\iota \kappa} F_{\lambda \mu}
+ \delta_{\iota \lambda} F_{\kappa \mu} +
\delta_{\iota \mu} F_{\kappa \lambda}
\nonumber\\ 
+\delta_{\kappa \lambda} F_{\iota \mu}
+\delta_{\kappa \lambda} F_{\iota \mu} + 
\delta_{\lambda \mu} F_{\iota \kappa} )
\nonumber\\ 
+(\delta_{\iota \kappa} \delta_{\lambda \mu} +
\delta_{\iota \lambda}\delta_{\kappa \mu} + 
\delta_{\iota \mu} \delta_{\kappa \lambda} ) ],
\label{e19}
\ee
where
\be
G_{\iota \kappa \lambda \mu}= 
\frac{1}{N} \sum_{j=1}^N 
\left(w_{j,\iota} w_{j,\kappa} w_{j,\lambda} w_{j,\mu}\right).
\ee
The corresponding 
frame-independent (rotationally invariant) order parameters
are defined by 
\be
 O_L = \sum_{j=1}^N \sum_{k=1}^N P_L (\mathbf{w}_j \cdot 
 \mathbf{w}_k) \ge 0,
\la{eqgenop01}
\ee
where the inequality follows from the addition theorem for spherical harmonics
\ci{ref00};
the order parameters of a configuration
are thus defined by
\be
\tau_L 
= \frac{1}{N}\ \sqrt{ O_L },
\ee
overall averages over the simulation chain are
\be
\overline{\tau}_L 
= \frac{1}{N}\langle \sqrt{ O_L } \rangle,
\ee
and the  associated susceptibilities  read
\be
\chi_L= \frac{1}{N} \beta
\left( \langle O_L \rangle- \langle \sqrt{ O_L } \rangle ^2 \right),  
\ee
where $\beta=1/T^*$.

Notice  that,
by the addition theorem for spherical harmonics
\ci{ref00},
 Eq. (\ref{eqgenop01}) can  actually be  calculated 
\ci{rcu03,rtetsim01,pre074011704} 
via  the   computationally more convenient 
single-particle sums
\be
\xi_{L,m}=\sum_{j=1}^{N}  \Re\left[Y_{L,m}(\mathbf{w}_j)\right],~
\eta_{L,m}=\sum_{j=1}^{N} \Im\left[Y_{L,m}(\mathbf{w}_j)\right];~
\la{eqadd01}
\ee
here $m=0,1,2,\ldots L$, $Y_{L,m}  (\ldots)$ are  spherical
harmonics, and $\Re$ and $\Im$ denote real and imaginary parts, respectively;
in turn, each spherical harmonic
is a suitable polynomial constructed in terms
of Cartesian components of the corresponding unit vector
(see, e.g. Ref. \ci{rLN}).

Moreover,
\be
O_2 = \frac{2}{3}\left( \mathsf{Q} : \mathsf{Q} \right) =  \frac{2}{3}
 \sum_{k=1}^3 q_k^2,
\\
O_4 = \frac{8}{35}\left( \mathsf{B} : \mathsf{B} \right).
\ee
There exist a  few different but related possible measures of second-rank order,
 i.e., in addition to $\tau_2$, one can consider the eigenvalues
$q_{max}$ or $q^{\prime}$,
with 
$O_2=|q^{\prime}|^2$ in the uniaxial case; $\tau_2$ and $q_{max}$ have a definite sign,
whereas the sign of $q^{\prime}$ may fluctuate in the course of simulation, and better take into account
configurations with antinematic order; for example, $D_2$-type configurations
(Appendix \ref{appA})
yield $q^{\prime}=-1/2$, $q_{max}=+1/4$, $\tau_2=+1/2$.
 
For  $L=4$, the above $\mathcal{T}^{(4)}$
tensor can be copied (``folded'') into a real, symmetric and traceless
matrix of order 9, say $\mathsf{H}$ \ci{rcu07,rrr001,rrr002}, where, for example,
\be
H_{\nu \rho} = B_{\iota \kappa \lambda \mu},~\nu = 3(\iota-1)+ \lambda,~\rho = 3(\kappa-1)+\mu,
\la{BandH}
\ee
and
\be
 \mathcal{T}^{(4)}:\mathcal{T}^{(4)} = \mathsf{H}:\mathsf{H}.
\la{BandHres}
\ee
The matrix $\mathsf{H}$ can be diagonalized to give the real eigenvalues
$\left\{\zeta_k,~k=1,2,\ldots,9\right\}$; thus
\be
O_4 = \frac{8}{35} \sum_{k=1}^9 \zeta_k^2.
\ee
Here also  there exist a  few different and related possible measures of fourth-rank order,
 i.e, in addition to $\tau_4$, one can consider the eigenvalue $\zeta^{\prime}$
with maximum absolute value, or the maximum eigenvalue $\zeta_{max}$.
Besides the above procedure
(Eq.  (\ref{BandH})), other computational definitions of fourth-rank 
orientational order are also possible \ci{rcu00}.

In principle, the same spin configuration might exhibit
(or the same underlying interaction model might produce) different types of 
ordering,  and the above definitions make it possible to calculate them
independently of one another, in contrast to  usual procedures
for order parameters in nematic liquid crystals, where the definition 
of $\overline{P}_4$ and higher-order terms is physically bound to the
director frame \ci{r23,r24,bpz-02}.
A few spin configurations possessing fourth-rank but no second-rank
orientational order are presented  in  Appendix \ref{appB}; for $D_3$-type configurations,
$\tau_4=\sqrt{21}/9 \approx 0.5092$

\newpage
\parindent 0cm
\section{Results}\label{results}
\begin{figure}[ht!]
\includegraphics[scale=0.4]{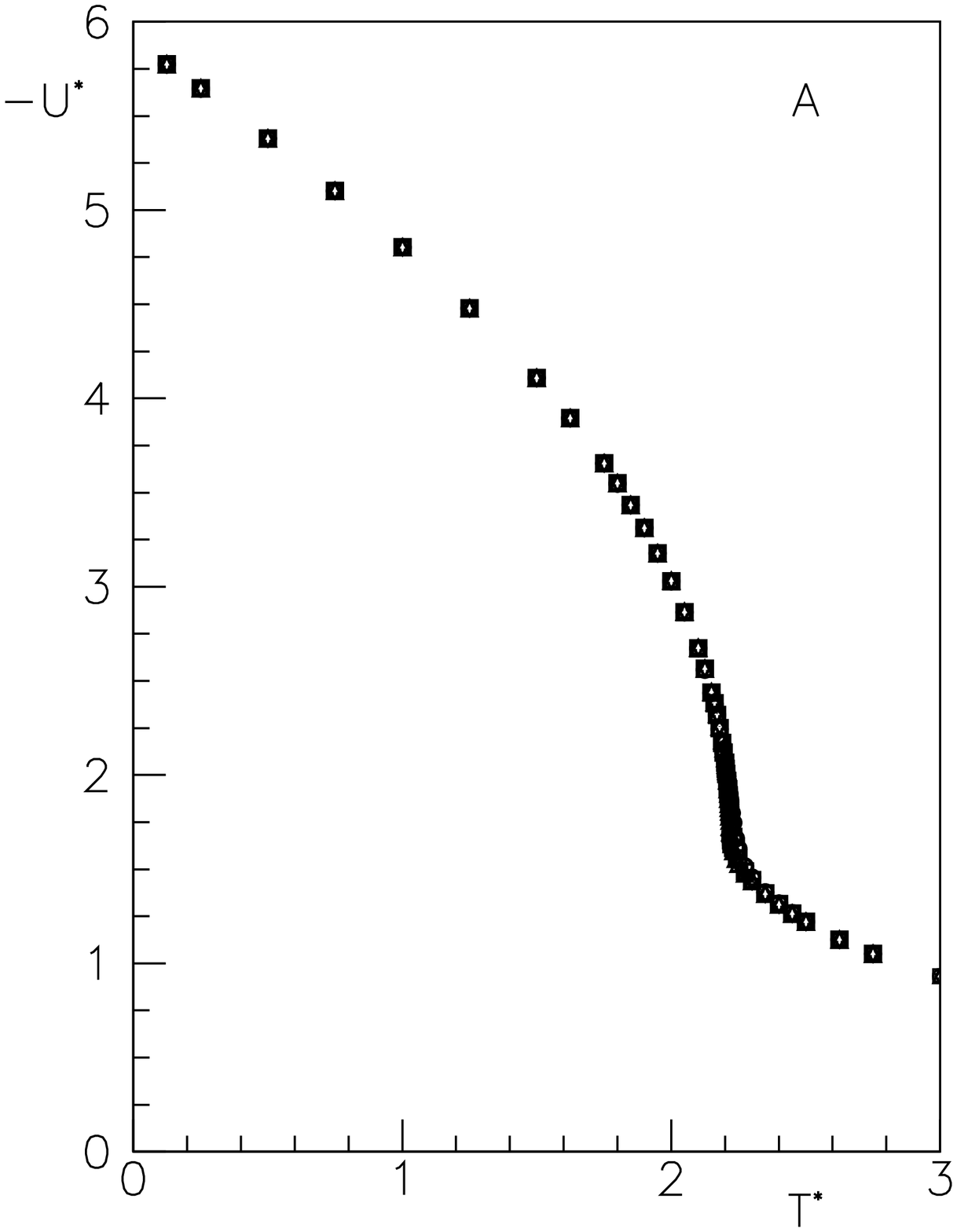}
\includegraphics[scale=0.4]{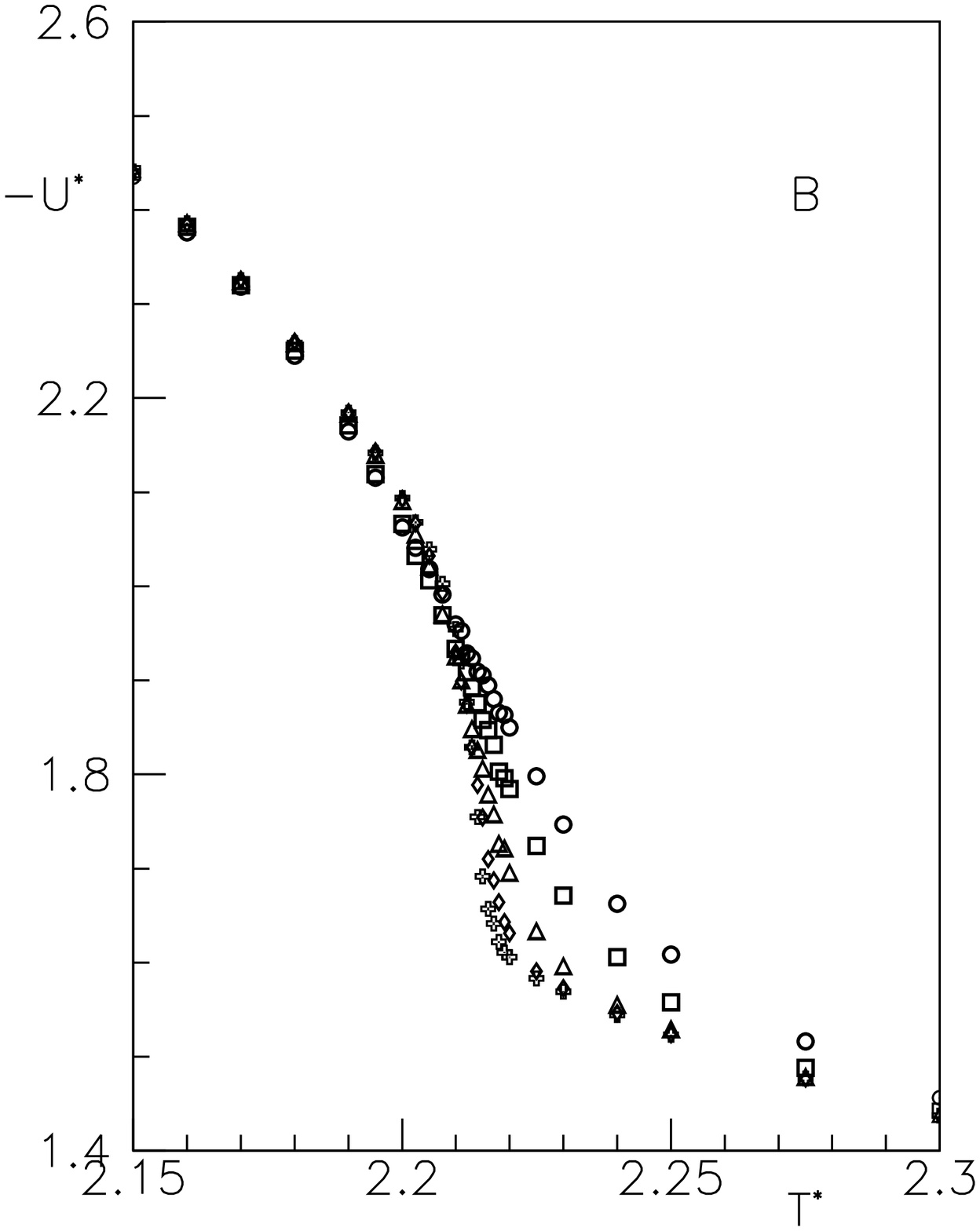}
\caption{Simulation results for the potential energy, obtained with different sample sizes.
Meaning of symbols: circles: $l=10$; squares: $l=12$; triangles: $l=16$;
diamonds: $l=20$, crosses: $l=24$.
Here and in the following figures,  with the exception of simulation results
for $C^*$, the statistical errors  mostly fall within symbol sizes.
Here as well as for Figs. (\ref{f02}), (\ref{f04}). and  (\ref{f05}), subfigure A covers the whole investigated temperature range, and subfigure B presents the transition region in greater detail.}
\label{f01}
\end{figure}

Simulation results for the potential energy $U^*$ (Figs. (\ref{f01})) 
appeared to exhibit a gradual monotonic change with temperature; they were found to be
independent of sample size up to $T^*_1 \approx 2.2$ and then above $T^*_2 \approx 2.3$;
their overall temperature behavior suggested 
a change of slope at some intermediate temperature $\approx 2.21$.
%
%
%

\begin{figure}[ht!]
\includegraphics[scale=0.4]{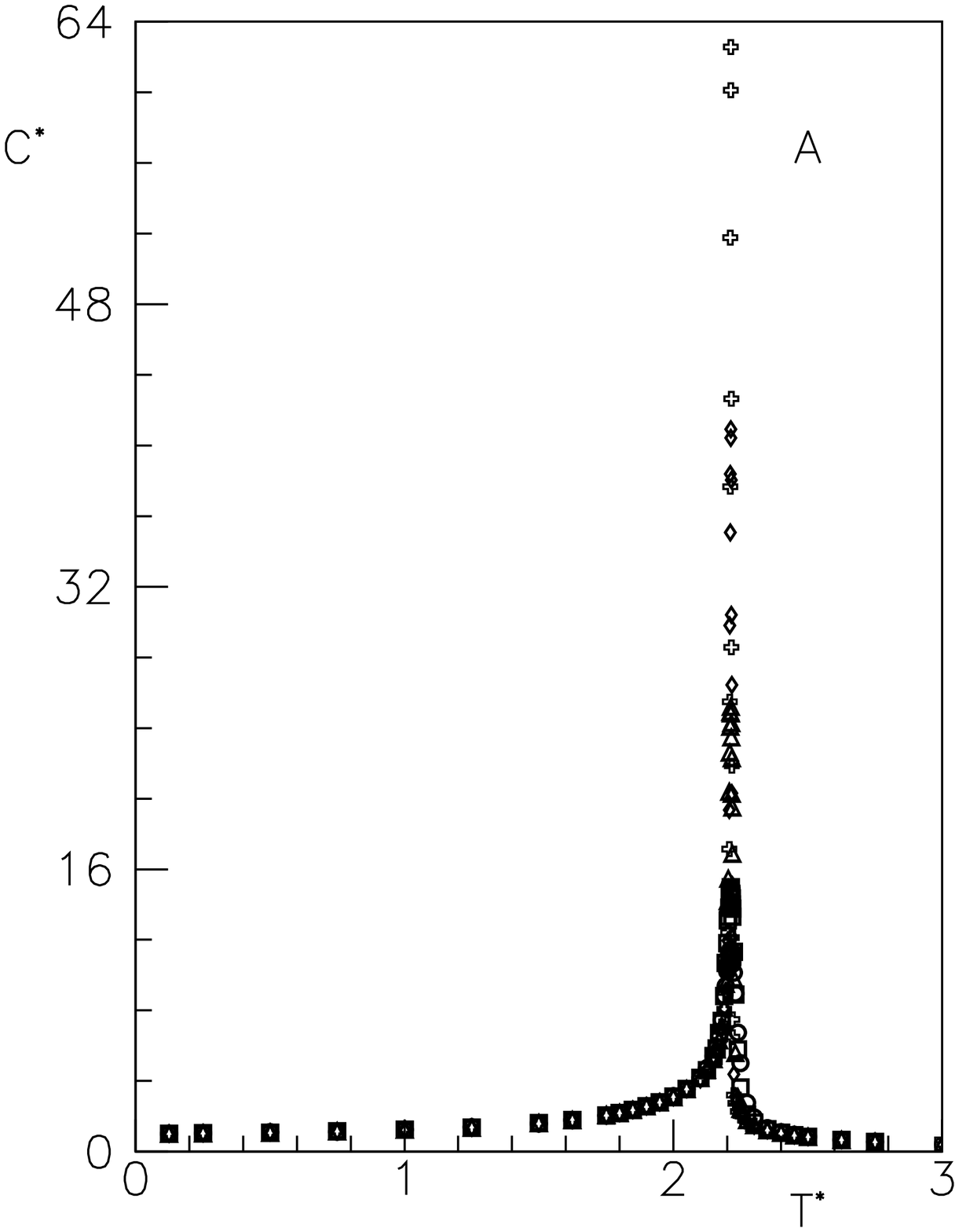}
\includegraphics[scale=0.4]{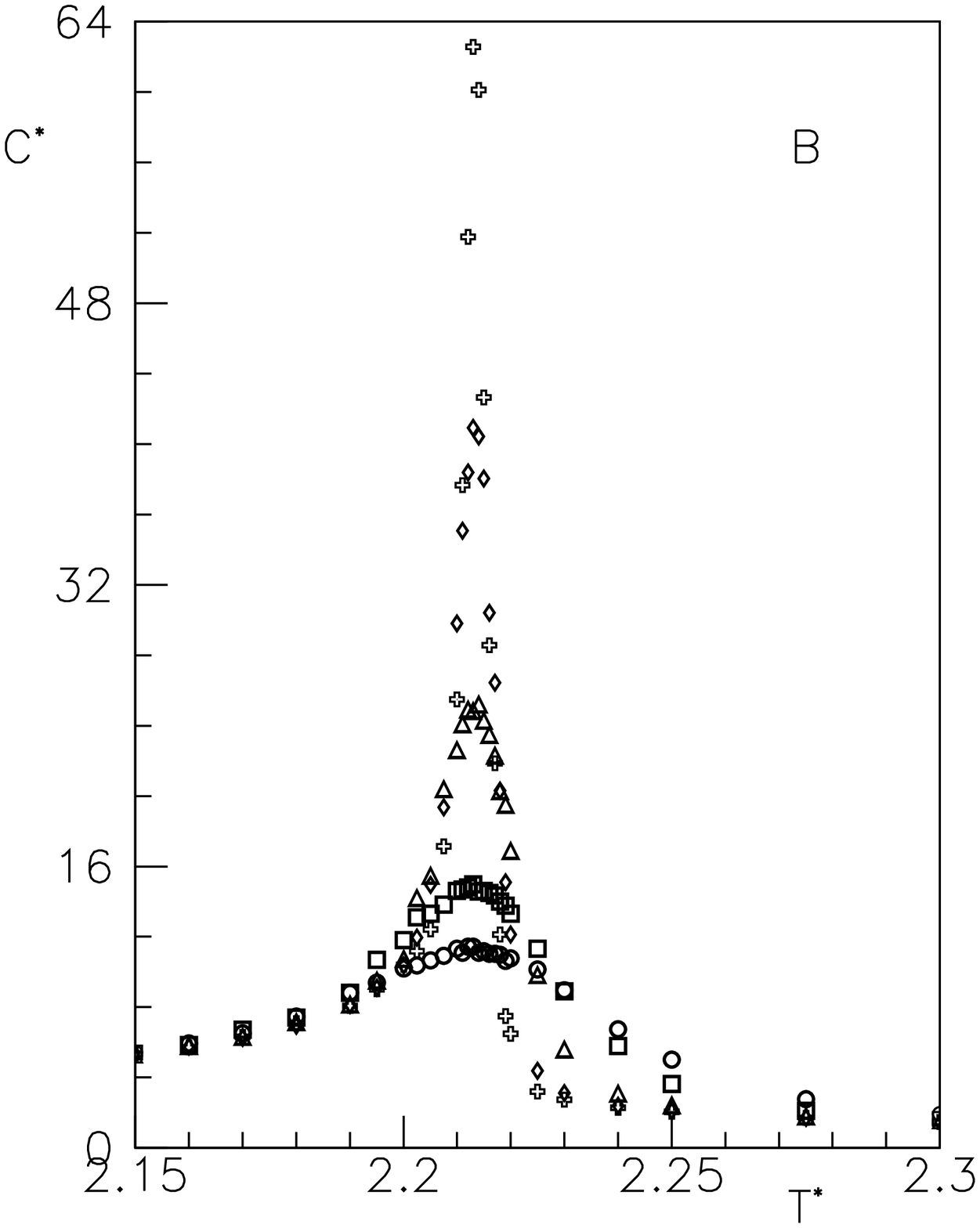}
\caption{Simulation results for the 
configurational specific heat $C^*$, obtained with different sample sizes;
same meaning of symbols as in Fig. (\ref{f01}).
The associated statistical errors, not shown, range between 1 and 5 \%.
}
\label{f02}
\end{figure}

\begin{figure}[ht!]
\includegraphics[scale=0.4]{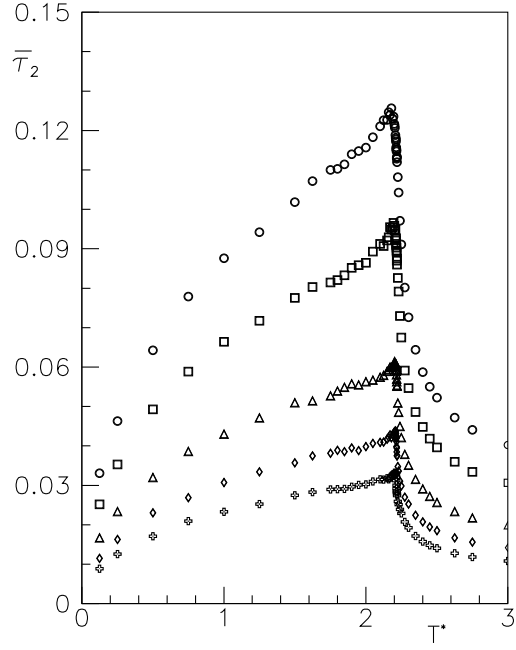}
\caption{Simulation results for the orientational order parameter
$\overline{\tau}_2$,
obtained with different sample sizes;
same meaning of symbols as in Fig. (\ref{f01}).}
\label{f03}
\end{figure}

The configurational specific heat $C^*$ (Figs. (\ref{f02})) 
was also found to be unaffected by sample size outside the named
temperature  range $[T^*_1,~T^*_2]$; in that range sample-size effects became
quite apparent, and a peak was found to develop at $\approx 2.21$,
growing narrower and higher with increasing sample size.
%
%
The second-rank order parameter $\overline{\tau}_2$ was calculated as well;
at all examined temperatures, the results were found to keep decreasing with increasing sample size
on the other hand, for all examined sample sizes, they were found to increase 
with temperature, reach a maximum in the transition region, and then
decrease with increasing temperature (Fig. (\ref{f03})).

~~~

\REM{ 

RA - 01

1) The decrease of the second order parameter as sample size increase
is convincing evidence of its eventual vanishing. However the finite
size samples show a strange temperature dependence with the order
decreasing at low temperature (fig.5). A comment on this is probably
needed.
 } 

\begin{figure}[ht!]
\includegraphics[scale=0.4]{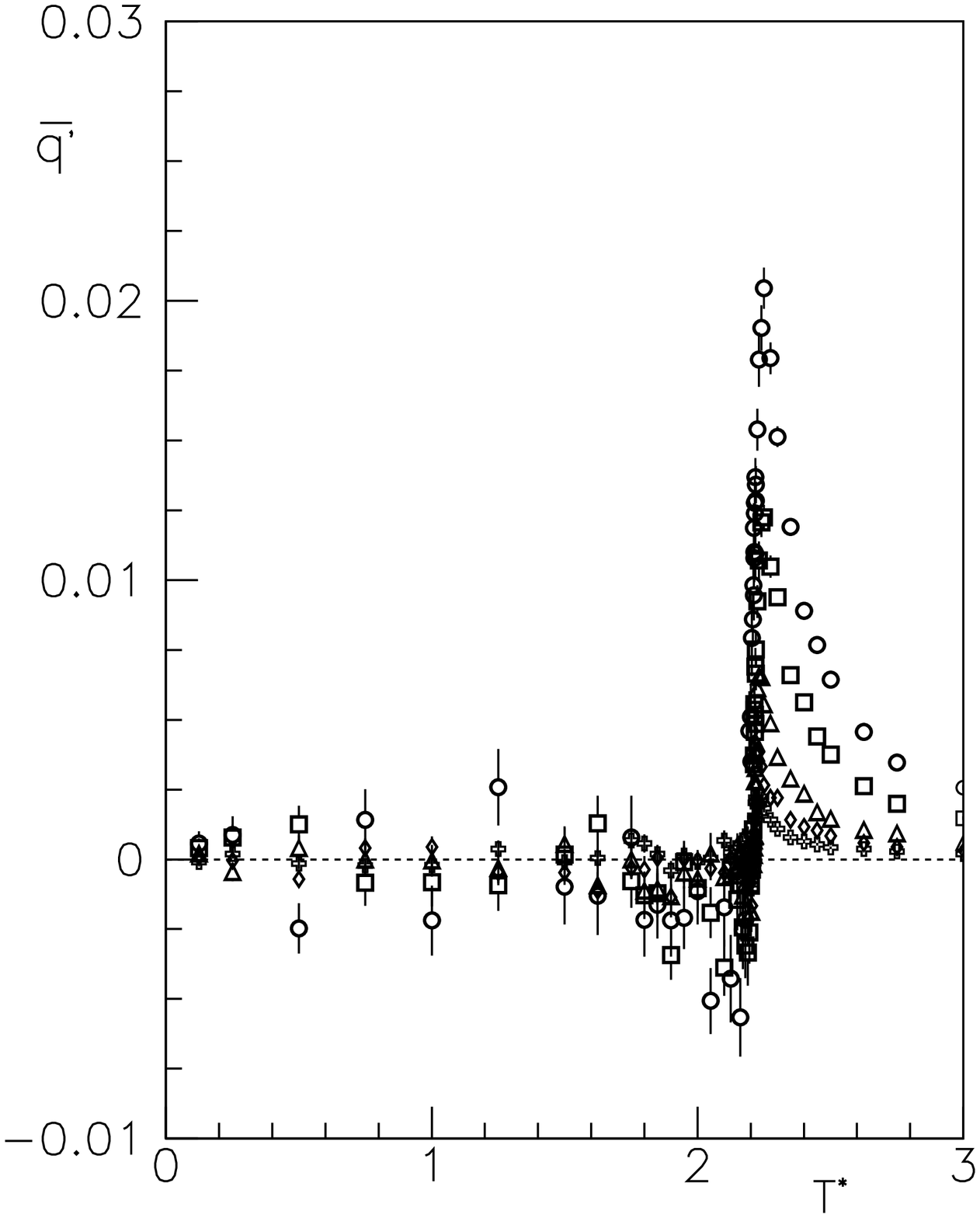}
\caption{Simulation results for the orientational order parameter
$\overline{q^{\prime}}$,
obtained with different sample sizes;
same meaning of symbols as in Fig. (\ref{f01}).}
\label{f03add}
\end{figure}

Notice that configurations possessing some
amount of second-rank orientational order 
may have potential energies not too high   above the ground state  of the 
model under investigation here
(Sect. \ref{ground} and Appendix \ref{appA});
in the low-temperature ordered and in the transition regions they can be  favoured by thermal
fluctuations; in turn, at a given temperature,  thermal fluctuations tend to be reduced
with increasing sample size.
On the other hand, 
as pointed out in the previous Section, there are different   possible computational
measures of second-rank order for a configuration;  
usage of $q^{\prime}$ yielded
absolute values smaller by roughly an order of magnitude, and sometimes
negative signs in the ordered region, as shown in Fig. (\ref{f03add}).
The corresponding susceptibility $\chi_2$ (not shown) 
was found to be less affected by sample size.

%
%
\begin{figure}[ht!]
\includegraphics[scale=0.4]{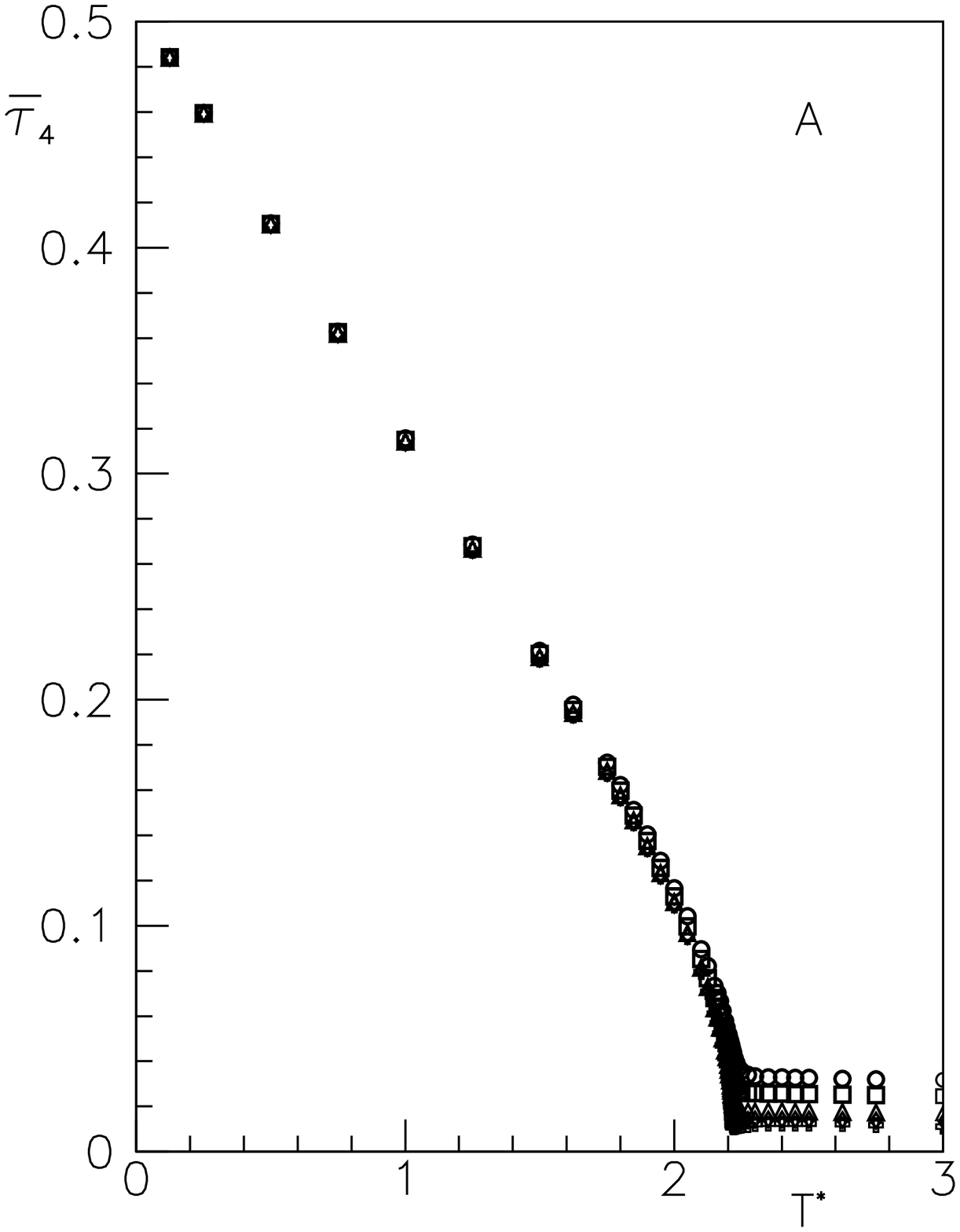}
\includegraphics[scale=0.4]{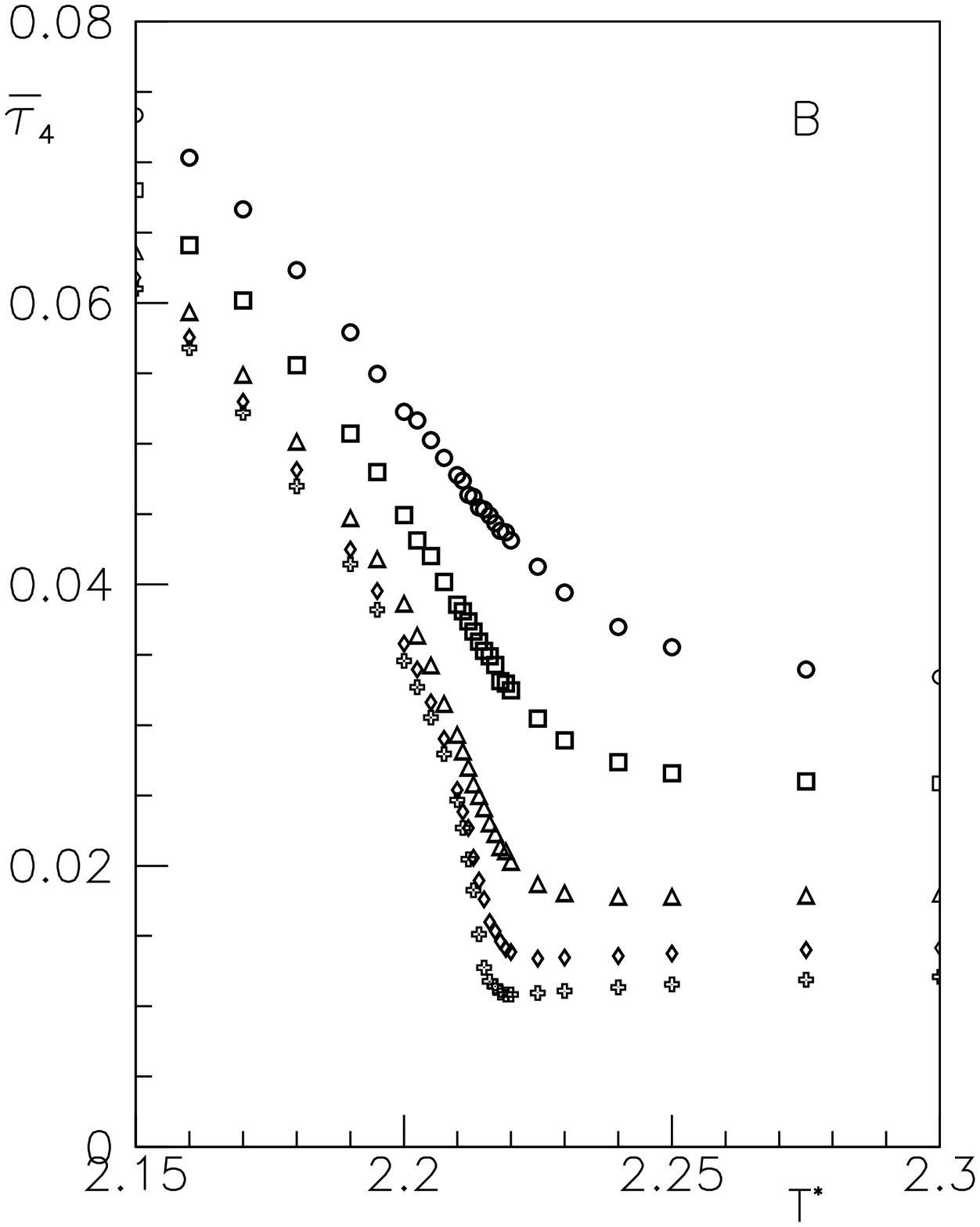}
\caption{Simulation results for the orientational order parameter
$\overline{\tau}_4$
obtained with different sample sizes;
same meaning of symbols as in Fig. (\ref{f01}).}
\label{f04} 
\end{figure}

Simulation results for the fourth-rank order parameter $\overline{\tau}_4$ were found to be independent 
of sample size up to $T^* \approx 2.1$, and then developed a recognizable decrease
with increasing sample size; their overall temperature behavior seemed to suggest
a continuous evolution with temperature (Figs. (\ref{f04})).  

\begin{figure}[ht!]
\includegraphics[scale=0.4]{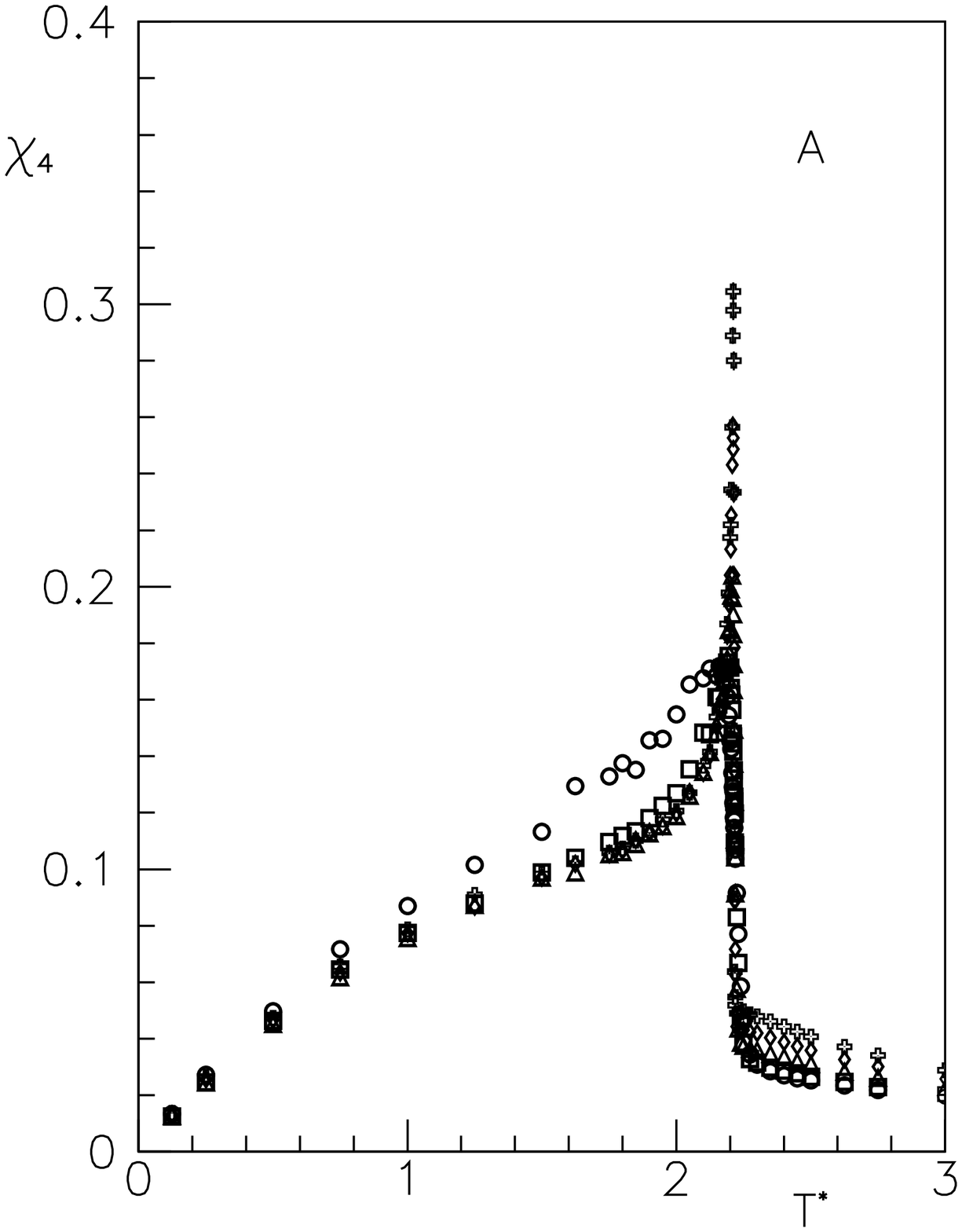}
\includegraphics[scale=0.4]{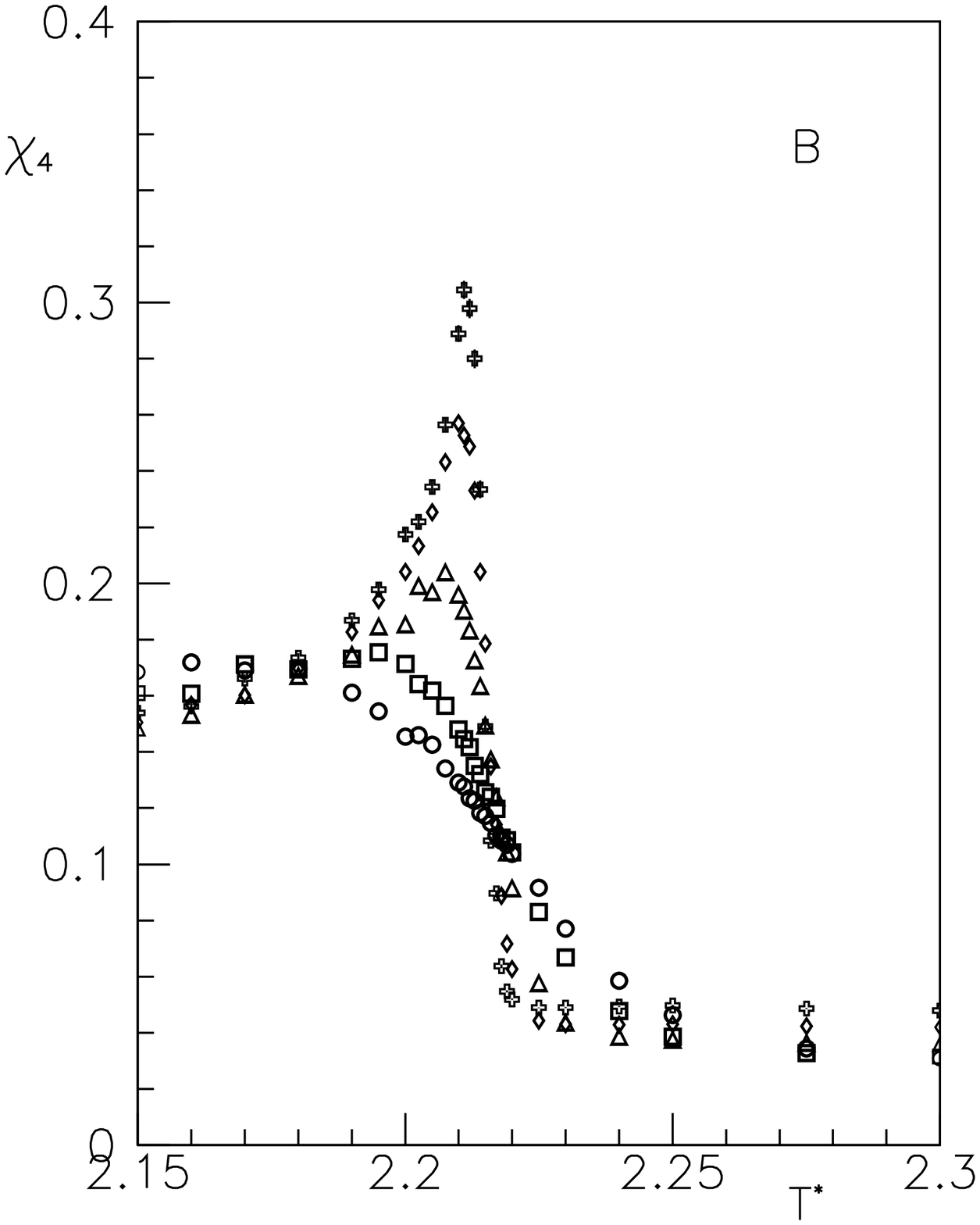}
\caption{Simulation results for the susceptibility $\chi_4$,
obtained with different sample sizes;
same meaning of symbols as in Fig. (\ref{f01}).}
\label{f05} 
\end{figure}

\begin{figure}[ht!]
\includegraphics[scale=0.4]{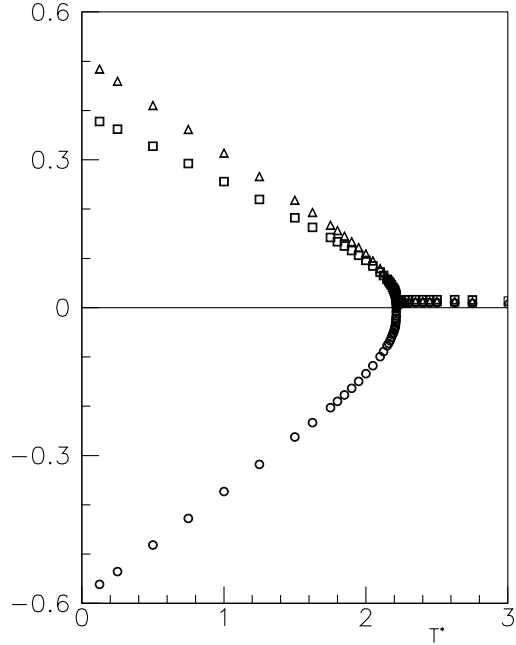}
\caption{Comparison between different  definitions
of the fourth-rank order parameter, based on simulation
results obtained for the largest investigated sample size $l=24$.
Meaning of symbols: circles $\overline{\zeta}^{\prime}$;
squares: $\overline{\zeta}_{max}$; triangles: $\overline{\tau}_4$.}
\label{f06} 
\end{figure}

Simulation results for the corresponding susceptibility $\chi_4$ were found to
increase with temperature (and to be unaffected by sample sizes for $l>10$)
up to the above transition region, where a peak developed, growing higher and
narrower with increasing sample size (Figs. (\ref{f05})). 
%
%
%
%
As pointed out in Sect. \ref{comptaspect},
 different measures of fourth-rank orientational
order can be defined, involving $\tau_4$, 
$\zeta^{\prime}$, or $\zeta_{max}$, respectively;
simulation results obtained for the three definitions and with the
largest investigated sample size $l=24$ are compared in Fig. (\ref{f06}),
where the three definitions appear to be mutually compatible; ratios 
between  pairs of them (at the same temperature) were also calculated, and found
to evolve slowly with temperature in the ordered region,
where they remained close to their ground-state values (see also  Appendix \ref{appB}).

\REM{ 


2) One of the main findings is that the observed transition to cubatic
phase is second order, but hardly any evidence or discussion of this
point is offered. Fig.2 and 3 show a clear increase of the specific
heat with size that could also indicate a first order transition. A
more convincing evidence of the character of the transition, e.g.
using histograms of energy values could be given (see, e.g., for the
Lebwohl-Lasher model, Fabbri & Zannoni, Mol. Phys. 58, 763 (1986) and
Chiccoli & al., Physica A, 148A, 298 (1988)).

} 

\begin{figure}[ht!]
\includegraphics[scale=0.4]{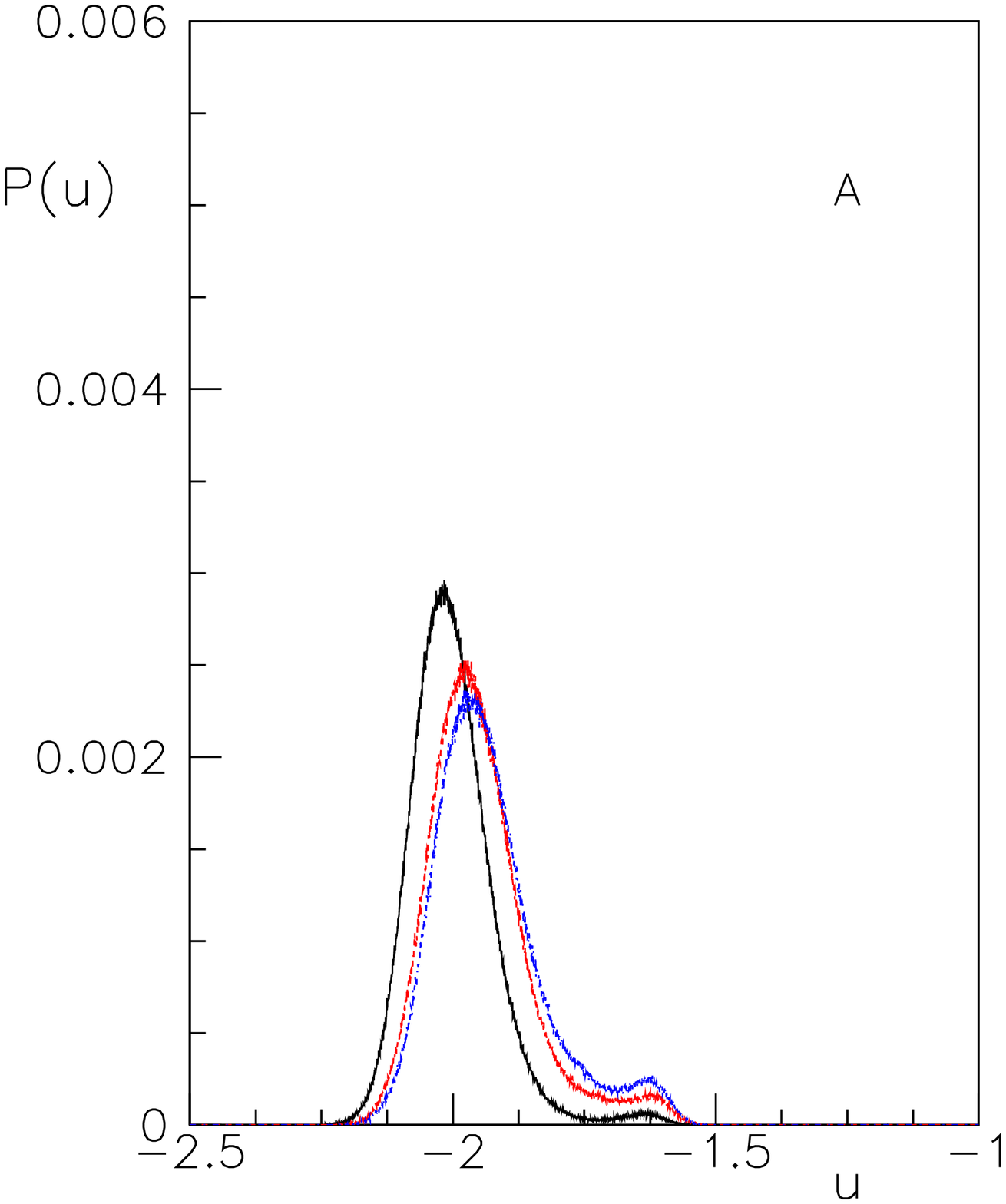}
\includegraphics[scale=0.4]{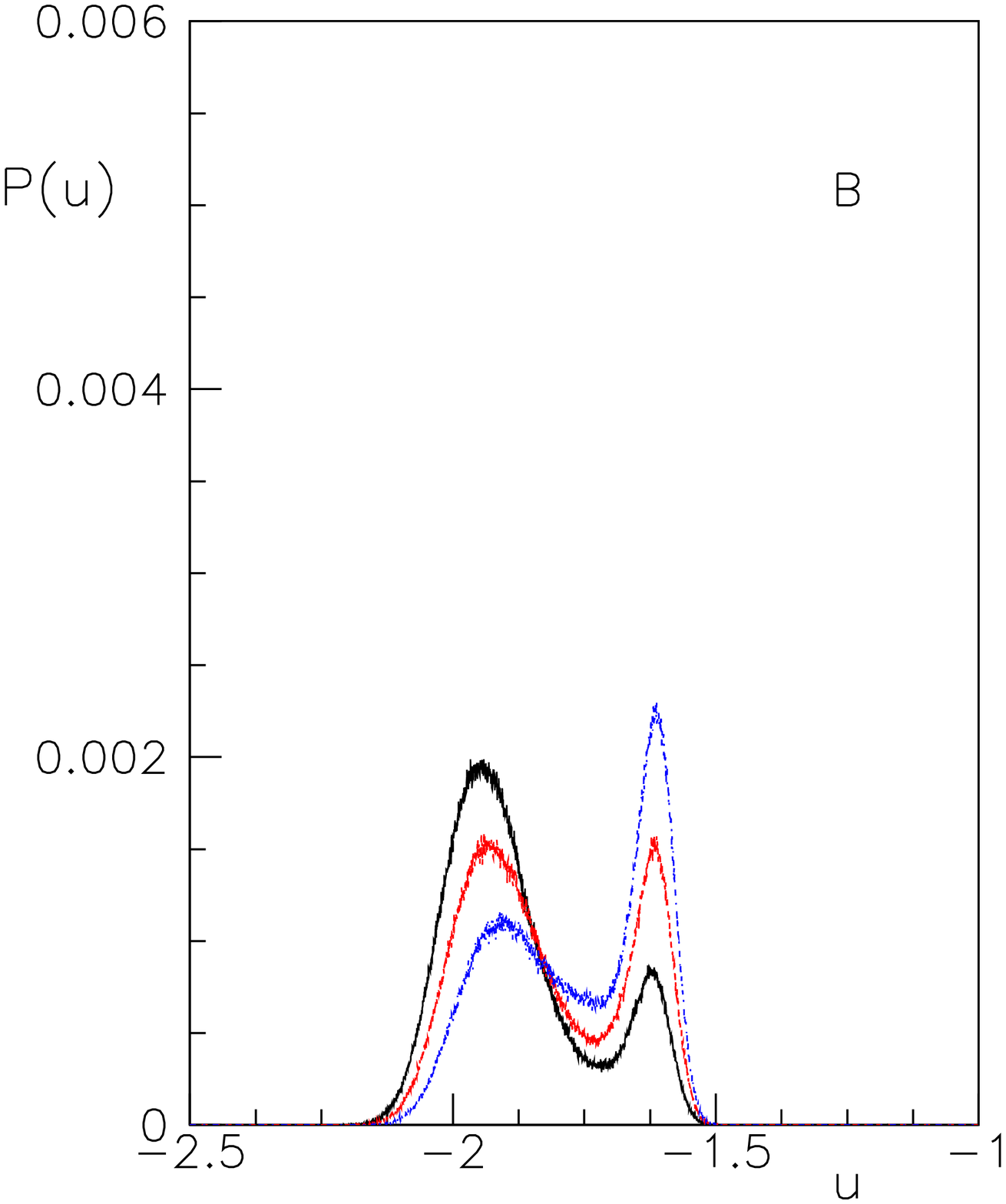}
\includegraphics[scale=0.4]{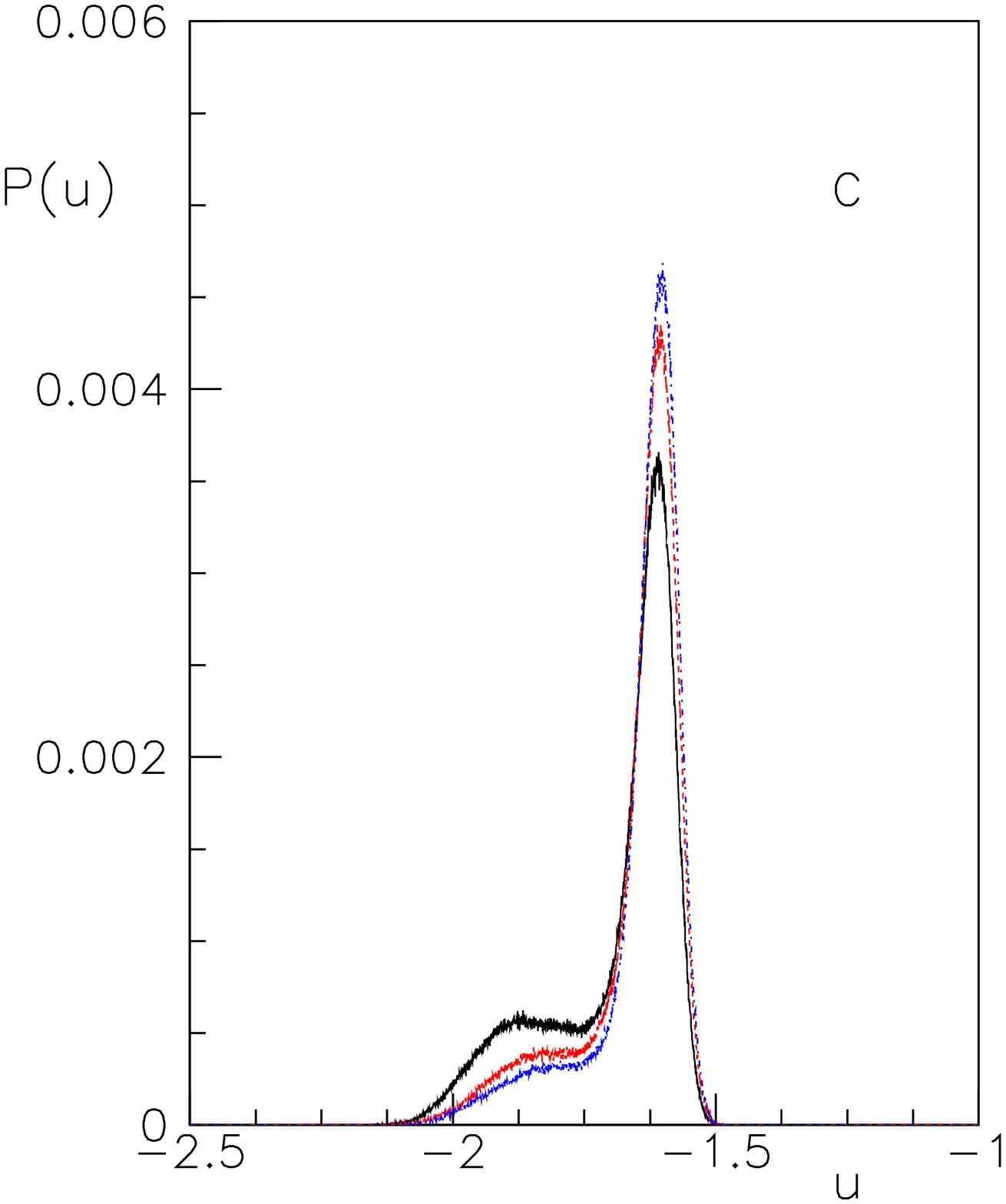}
\caption{(Color on line) Histograms  of the single-particle potential energy $u$, obtained for $l=24$ and different temperatures in the transiton region.
Meaning of symbols for subfigure A: black continuous line: $T^*=2.2075$;
red dashed line: $T^*=2.210$; blue dashed-dotted line: $T^*=2.211$.
Meaning of symbols for subfigure B: black continuous line: $T^*=2.212$;
red dashed line: $T^*=2.213$; blue dashed-dotted line: $T^*=2.214$.
Meaning of symbols for subfigure C: black continuous line: $T^*=2.215$;
red dashed line: $T^*=2.216$; blue dashed-dotted line: $T^*=2.217$.}
\label{f07} 
\end{figure}

\begin{figure}[ht!]
\includegraphics[scale=0.4]{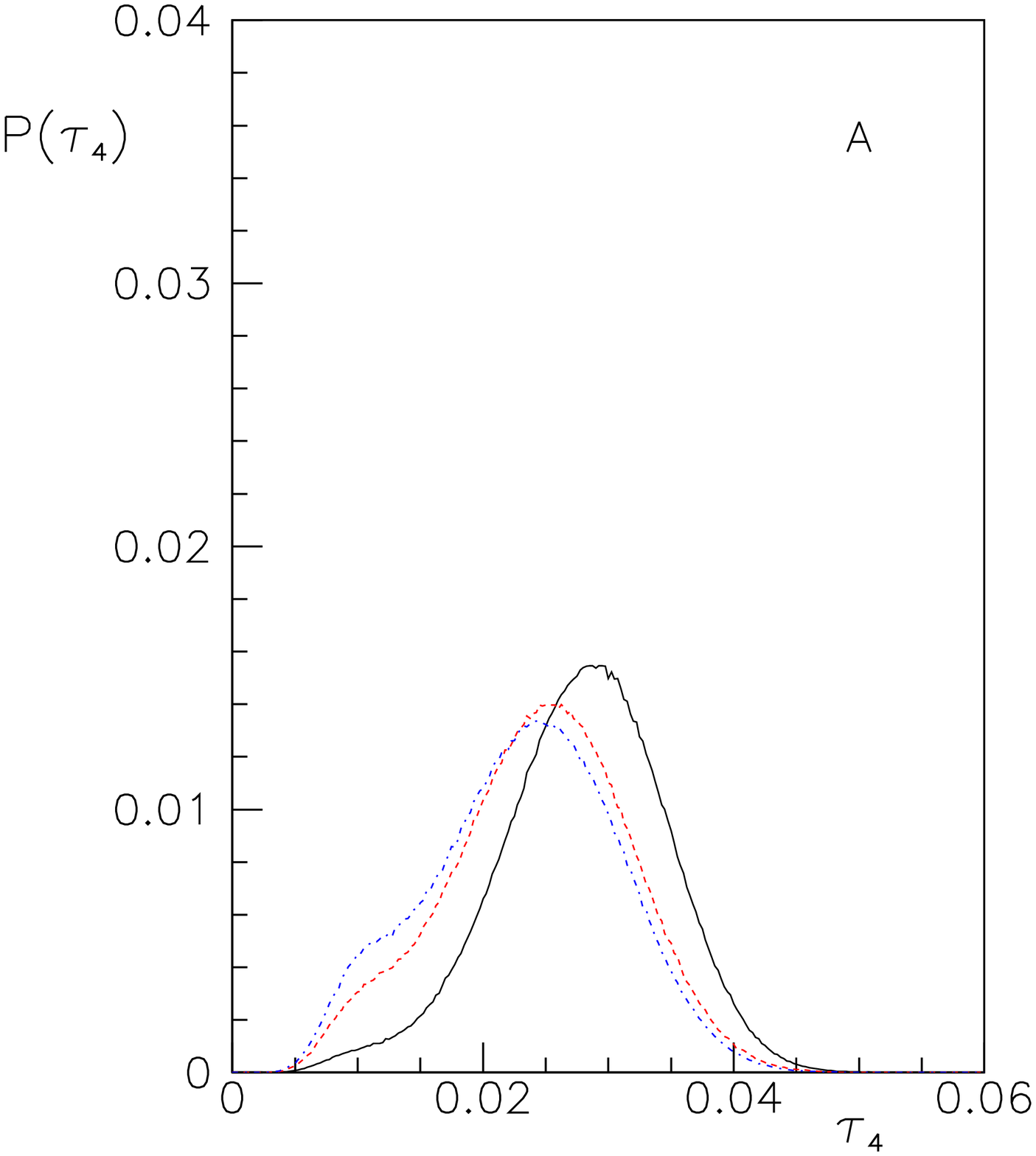}
\includegraphics[scale=0.4]{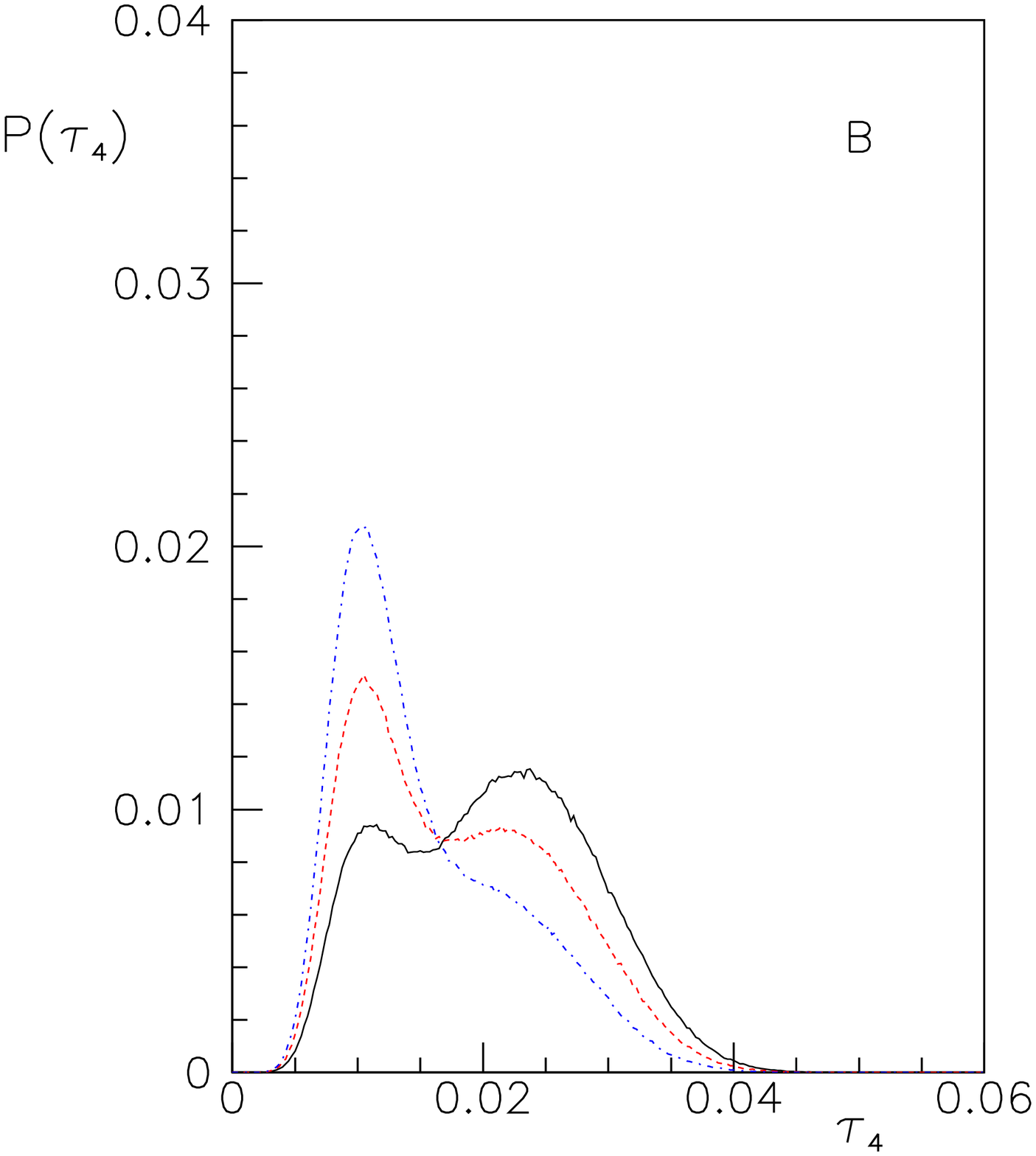}
\includegraphics[scale=0.4]{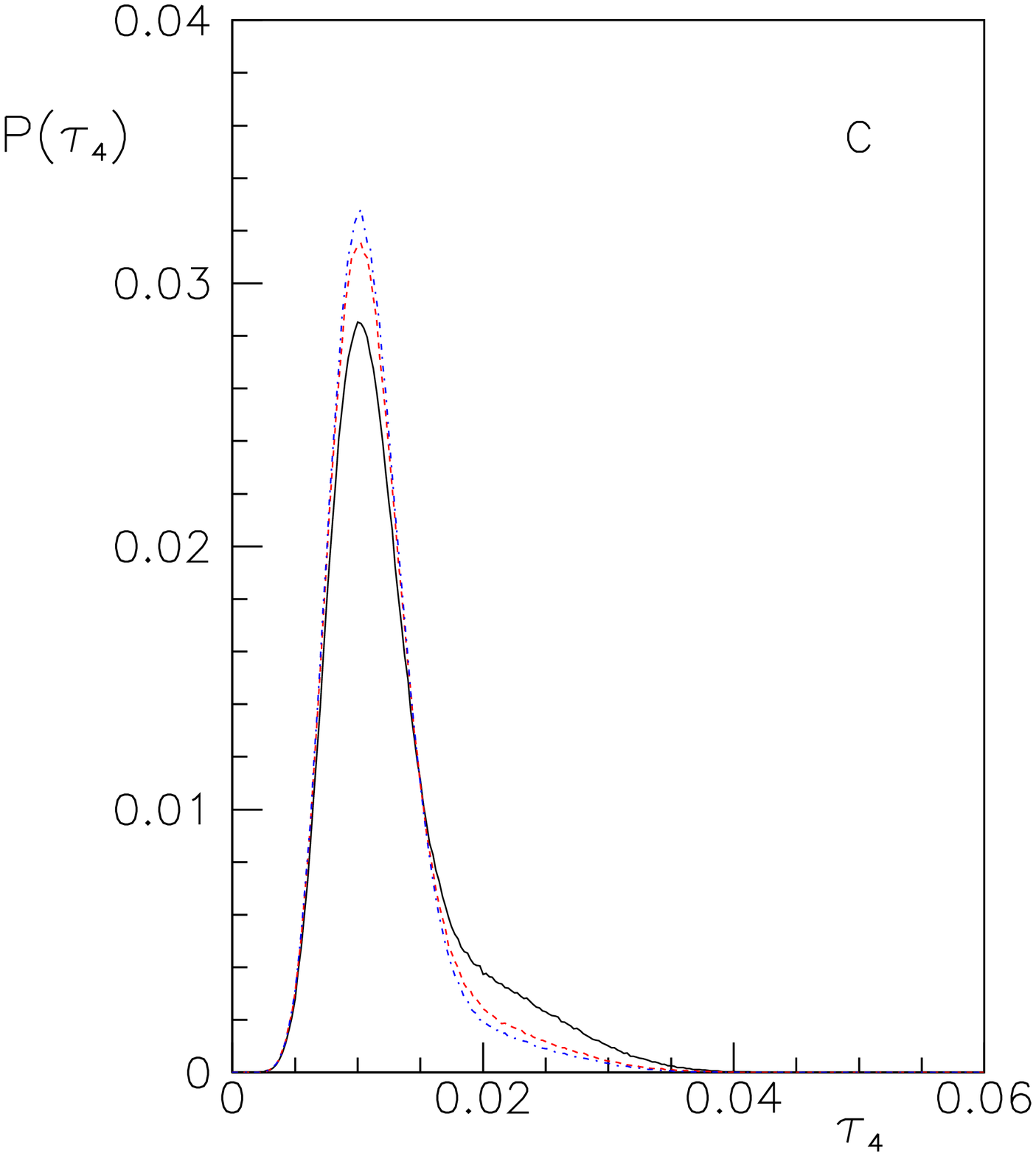}
\caption{(Color on line) Histograms  of the fourth-rank order parameter $\tau_4$,
obtained for $l=24$ and  different temperatures in the transiton region.
Meaning of symbols for subfigure A: black continuous line: $T^*=2.2075$;
red dashed line: $T^*=2.210$; blue dashed-dotted line: $T^*=2.211$.
Meaning of symbols for subfigure B: black continuous line: $T^*=2.212$;
red dashed line: $T^*=2.213$; blue dashed-dotted line: $T^*=2.214$.
Meaning of symbols for subfigure C: black continuous line: $T^*=2.215$;
red dashed line: $T^*=2.216$; blue dashed-dotted line: $T^*=2.217$.}
\label{f08} 
\end{figure}

 The above results point to a transition between cubatic and 
isotropic phase, taking place at $T^* \approx 2.213$;
 in order to obtain some more evidence of its
thermodynamic character, histograms for the frequency distribuition
$P(u)$ (where $u$ denotes the scaled 
 potential energy per particle, and  $U^*=\langle u \rangle$) as well as  for $P(\tau_4)$ 
(see, {\it e.g.}, Refs. \ci{rrr003,rrr004,rrr005})
were calculated in the transition region ($T^*=2.2075,~2.210$ to $2.220$ with step $0.001$,
$T^*=2.225$), 
for all examined sample sizes, over
additional  run lengths
ranging between
$1\,000\,000$ and $1\,500\,000$ cycles, and by analyzing one configuration every cycle.
Results for $P(u)$ and $l=24$
at selected temperatures  are plotted  in Fig. (\ref{f07}), 
and their counterparts for $P(\tau_4)$ are reported in Fig. (\ref{f08}).
The width of the distribution $P(u)$ as measured by the variance (not shown)
was found to shrink with increasing sample size;
the double-peaked structure in Fig, (\ref{f07}-B) only developed for
$l \ge 16$, and the peaks appeared to grow higher and narrower
with incresing sample size.
As for $P(\tau_4)$, the width of the distributionm was found to shrink with 
increasing sample size as well, and the double-peaked structure in Fig. (\ref{f08}-B)
only developed for $l \ge 20$.
In both cases, histograms obtained at higher temperatures, not shown,  exhibited rather narrow single peaks;
thus histograms obtained for large samples exhibit a 
two-peak structure over a rather narrow temperature range, pointing
to a weak
first-order transition taking place at $T \approx 2.213$.

%

%
\parindent 0cm
\section{Conclusions} \la{concl}
We have carried out a MC simulation study of a lattice model consisting
of uniaxial ($D_{\infty h}-$symmetric) particles coupled by long-range
dispersion interactions of the $LHB$ type; in the named setting the
model becomes equivalent to its NS counterpart; 
the model was found
to support no second-rank order but to possess fourth-rank order
in its low-temperature phase; 
simulation results point to a weak first-order transition
 whose transition temperature is  estimated
to be $T^*_c =2.213 \pm 0.002$, where  the uncertainty is conservatively
taken to be twice the temperature step used in the transition region.
Let us recall that the n-n counterpart supports \ci{rdisp1,rHR}
a first-order transition to a nematically ordered phase, taking place
at $2.238 \pm 0.001$;  the n-n model
is reasonably well described by a MF treatment of the MS type,
predicting a transition temperature $2.6424$;
as already pointed out in Ref. \ci{rdisp2}, a MF treatment of the MS type
could  be applied in this case as well: it would produce the same pseudopotential as in the n-n case (within a scaling factor), but here it would be 
physically wrong.

The investigated potential model involves second-rank interaction terms and produces
no second-rank but only
fourth-rank orientational order; it may be appropriate to mention that a few
other potential  models are known 
in the Literature, where  interactions of a certain rank  produce
 only higher-rank
correlations or even long-range orientational
order: more explicitly,  they  are  classical lattice-spin models 
involving $2$- or $3$-component unit vectors coupled by 
competing and frustrating  first-rank (magnetic) interactions, which may result
in  second-rank correlations or even long-range order at finite temperature
\ci{rrr006,rrr007,rrr008};
this appears to happen  via a mechanism of entropic selection (order by disorder)
of ground-state configurations, in contrast to
the energetic effects acting here (Sect. \ref{ground}).

There are also a few  related interaction models involving uniaxial particles 
and which 
seem to be worth examining or revisiting in terms of cubatic orientational 
order.
On the one hand, one can  study the effect of lattice geometry, by addressing 
BCC and FCC counterparts of the present simple-cubic lattice model;
one could even give up the lattice,  allow particle centers to move in 
$\mathbb{R}^3$,
 and supplement the present interaction 
model with a  purely radial term enforcing some  minimum distance between 
particle centers
(the ``liquid'' setting. for short);
there is a continuum counterpart of Eq. (\ref{eqiddisc}), i.e. 
\be
\int P_2\left(\mathbf{w} \cdot  \mathbf{v}\right) d^2 \mathbf{v} = 0,
\la{eqidcont}
\ee
where the unit vector $\mathbf{w}$ is assigned and  integration
over the unit vector $\mathbf{v}$ is carried out on 
the whole unit sphere with the usual uniform measure: 
this formula suggests that, in the ``liquid'' setting, the 
$S_{jk}$ terms (see Section \ref{ground})  should largely  cancel out
upon summing over all  interacting pairs, provided that the distribution of 
intermolecular vectors is essentially isotropic.

Interaction models involving  just linear point quadrupoles
associated with a $3-$d lattice were studied some forty  years ago,
analytically  \ci{rq01,rq02,rq03,rq04}
and by classical simulations \ci{rq05,rq06}, as simplified models of solid 
nitrogen \ci{rq07}, and might now be revisited, also  in terms of overall 
fourth-rank orientational order.
For example, the low-temperature, room-pressure $\alpha$ phase of solid 
nitrogen is usually assumed to belong to space group  $Pa3$ (but 
 space group $P2_{1}3$ is also possible  \ci{rq07,rq08}).
The $Pa3$ structure involves particle centers associated with a FCC lattice,
where the four particles in the cubic unit cell
are oriented  along the  body diagonals, thus producing   no overall 
second-rank orientational order, and  a finite amount of its fourth-rank 
counterpart, actually the same ground-state value $\tau_4=\sqrt{21}/9$  
as in our case (see also Appendix \ref{appB}); in  some original simulation 
papers a local (sublattice-wise) second-rank
order parameter was defined with respect to the corresponding ground-state 
orientations \ci{rq05,rq06}.

We hope to address some of these points in the near future.







\parindent 0cm
\section*{Acknowledgements}
The present extensive calculations were carried out, on,
among other machines,
workstations, belonging to the Sezione di Pavia of
Istituto Nazionale di Fisica Nucleare (INFN).
The author thanks Prof. J. T. Chalker (Oxford University, UK) for kindly
directing him to Refs. \ci{rrr007,rrr008}.

\newpage
\parindent 0cm
\appendix
\section{Ground state configuration} \la{appA} 
\REM{ 
A few  spin configurations possessing periodicity 2 in each lattice  directions 
and constructed as in Ref.  \ci{rdisp2}
were examined as possible  ground state candidates,
 and the results further 
checked by simulations carried out  at low temperatures.
} 
Let $(X,Y,Z)$ denote the Cartesian components of a unit vector,
parameterized by two polar angles $\left(\Theta,~\Phi\right)$;
for the generic   lattice site $\mathbf{x}_j$ let
$h,~k,~l$ denote the integer coordinate
(we write $h$ instead of $h_j$ for simplicity of notation), 
then, for each site $j$ \ci{rdisp2}.
\be
\mathbf{w}_j = (-1)^{k+l}  X \mathbf{e}_1 + (-1)^{h+l}  Y \mathbf{e}_2 + (-1)^{h+k}  Z \mathbf{e}_3.
\la{eqdipconf}
\ee
A few configurations defined by special cases of Eq. (\ref{eqdipconf}) are

 - $D_1$ ($X=Y=0,~Z=1$), with full orientational
   order along a lattice axis;

 - $D_2$ ($X=Y=\sqrt{2}/2,~ Z=0$), with  a negative
   second-rank order parameter $(-1/2)$ (antinematic order); 

 - $D_3$ ($X=Y=Z=\sqrt{3}/3$), possessing  no second-rank but a finite amount
   of fourth-rank order . 

\REM{ 
actually,  
Eq. (\ref{eqdipconf}) already entails that
each type of configuration may   be realized in different ways, for example by 
permuting components with different values,
and the even character of the interaction
further  increases the degeneracy.
} 

Let $W_1^*,~W_2^*,~W_3^*$ denote the corresponding
 potential energies per particle, where the asterisk means
scaling by $\epsilon$; known results 
when the interaction was truncated at n-n separation
\ci{rdisp1,rdisp2,rHR} are
\be
W_1^* = -6 < W_2^*=-21/4 < ~W_3^*=-5;
\la{grene01}
\ee
on the other hand, when the interaction was extended  to next-nearest or more
distant neighbors, the sequence became \ci{rdisp2}
\be
W_3^* < W_2^* < W_1^*,
\la{grene02}
\ee
and  the three values were found to be much closer to one another;
notice also that $D_2$-type configurations become  favoured over $D_1$-type.
Different truncation radii were tried, and found 
 to slightly change
the three individual values, but not the inequalities among them;
for example, 
truncation by the condition $\mathbf{r} \cdot \mathbf{r} \le 25$
yielded
\be
W_1^* = -5.813,~W_2^* = -5.879,~W_3^* = -5.900;
\la{grene03}
\ee
the configuration potential energy was also calculated over a finer angular
grid in  $\left(\Theta,~\Phi\right)$, and results appeared to confirm $D_3$
as ground-state candidate; moreover, and more importantly,   
 simulations carried out at low temperatures,
starting from any of the three above configurations, or even
for a randomly generated one, quickly gave results
corresponding to a mild thermal evolution of $D_3$, as originally found by simulation
in Ref. \ci{PSthesis}.


\section{Simple spin configurations possessing fourth-rank but no second-rank order} \la{appB}
On can construct a few simple spin configurations,
possessing,
fourth-rank (cubatic) order but no second-rank (nematic) one,  and for which
all odd-rank ordering tensors vanish;
\begin{enumerate}
\item a first example involves 6 unit vectors oriented along
$\pm \mathbf{e}_{\iota},~\iota=1,2,3$;
\item a second example consists of   8  unit vectors with Cartesian
components $\pm \sqrt{3}/3$ (all combinations of signs), and corresponds to the
ground state for the interaction model under investigation here; actually,
this configuration consists of two disjoint 
subsets, composed of  four spins whose components have an even number of
negative signs, and four spins with an odd number of negative signs, 
respectively; for both subsets and for
the whole configuration, $\tau_1=\tau_2=0,~\tau_4=\sqrt{21}/9$;
moreover, for each subset, $\tau_3=\sqrt{5}/3 \approx 0.7454$;
\item a  third example involves 12 unit vectors with Cartesian components obtained from $\left( 0,~ \pm \sqrt{2}/2,~\pm \sqrt{2}/2 \right)$ by applying all possible combinations of signs and all possible permutations; 
\item one can build a 26 spin configuration as union of the three above cases.
\REM{ 
\item a general configuration consisting of  48 spin can be constructed by taking 
three real positive spin components  $\left(X,~ Y,~ Z\right),~0<X<Y<Z$;
one can then start  from the Cartesian
components $\left( \pm X,~\pm Y,~\pm Z \right)$ and apply all  possible 
combinations of signs and all possible permutations; in turn, $X$, $Y$ and $Z$
 can be expressed in terms of two
polar angles $\left( \Theta,~\Phi \right)$, so that  $O_4$ is obtained 
  in closed form as a  trigonometric polynomial.
} 
\end{enumerate}

In all four cases the above matrix $\mathsf{H}$ (Eq. (\ref{BandH}))
was found to possess the  eigenvalue 0 with degeneracy 4,
as well as two  other nonzero ones, with opposite
signs ($\zeta_{-}$ and $\zeta_{+}$ in the following), degeneracies 2 or 3, respectively, and absolute values
in the corresponding ratio  (the eigenvalue with smaller magnitude possessing
higher degeneracy); moreover 
we found $\zeta^{\prime}>0$ for the first case, and $\zeta^{\prime}<0$
in all others; thus, in all the four cases, 

$\tau_4 = \sqrt{ (8/35) \sum_{k=1}^9 \zeta_k^2},~
\sum_{k=1}^9 \zeta_k^2 = (10/3) |\zeta^{\prime}|^2,~~
\tau_{4}/|\zeta^{\prime}| = (4/21) \sqrt{21} \approx 0.8728$.

\begin{table}[ht!]
\caption[]{Eigenvalues of  $\mathsf{H}$ (Eq. (\ref{BandH})) for the four
discussed spin configurations.}
\la{tconfigs}
\begin{tabular}{lccllcl}
case  & number of spins &~ & $\zeta_{-}$ & $\zeta_{+}$ &~& $\tau_4$
\\
\hline
~ & ~ & ~ & ~ & ~ & ~ & ~
\\
1 & 6 & ~
 & $-7/12 \approx -0.5833$ & $+7/8=+0.875$  &~& $\sqrt{21}/6 \approx 0.7638$  \\
~ & ~ & ~ & ~ & ~ & ~ & ~
\\
2 & 8 & ~ 
 & $-7/12 \approx -0.5833 $ & $+7/18 \approx +0.3889$ & ~ & $\sqrt{21}/9 \approx 0.5092$  \\
~ & ~ & ~ & ~ & ~ & ~ & ~
\\
3 & 12 & ~
 & $-7/32 = -0.21875$ & $+7/48 \approx +0.14583$  & ~ & $\sqrt{21}/24 \approx 0.1909$  \\
~ & ~ & ~ & ~ & ~ & ~ & ~
\\ 
4 & 26 & ~
 & $-49/624 \approx -0.0785$ & $+49/936 \approx +0.0523$ & ~  & $7\sqrt{21}/468 \approx 0.0685$  \\
~ & ~ & ~ & ~ & ~ & ~ & ~ 
\end{tabular}
\end{table}

~~~


\newpage
\begin{thebibliography}{999}%
\bi{rGRLnature}
G. R. Luckhurst, Nature {\bf 430}, 413 (2004).

\bi{book00} {\it Biaxial Nematic Liquid Crystals Theory, Simulation, 
and Experiment}, edited by G. R. Luckhurst and T. J. Sluckin (Wiley, 2015).

\bi{prl02400359}
T. R. Taylor, J. L. Fergason. and S. L. Arora,
Phys. Rev. Lett. {\bf 24}, 359 (1970). 

\bi{prl02401041}
M. J. Freiser,
Phys. Rev. Lett. {\bf 24}, 1041 (1970).
\bi{book15} G. R. Luckhurst, and T. J. Sluckin, Chap. 15 in Ref. \ci{book00}.


\bi{book14} M. Lehmann, Chap. 14 in Ref. \ci{book00}.


\bi{npb26500647}
M. V. Jari\'c, Nucl. Phys. B {\bf 265}, 647 (1986). 

\bi{pre05200702} L. G. Fel, Phys. Rev. E {\bf 52}, 702 (1995).

\bi{pre05202692} L. G. Fel, Phys. Rev. E {\bf 52}, 2692 (1995).

\bi{pre074041701}
B. Mettout, Phys. Rev. E {\bf 74}, 041701 (2006).

\bi{pre094022701} J. Nissinen, K Liu, R.-J. Slager, K. Wu, and J. Zaanen,
Phys. Rev. E {\bf 94}, 022701 (2016).

\bi{rev001} S. Torquato, and F. H. Stillinger, Rev. Mod. Phys. {\bf 82}, 2633 (2010).
\bi{rev002} U. Agarwal, and F. A. Escobedo, Nature Mater. {\bf 10}, 230 (2011).
\bi{rev003} J. de Graaf, R. van Roij, and M. Dijkstra, Phys. Rev. Lett. {\bf 107}, 155501 (2011).
\bi{rev004} S. Torquato, and Y. Jiao, Phys. Rev. E {\bf 86}, 011102 (2012).
\bi{rev005} P. F. Damasceno, M. Engel, and S. C. Glotzer, Science {\bf 337}, 453 (2012).
\bi{rev006} D. Chen, Y. Jiao, and S. Torquato,
J. Phys. Chem. B {\bf 118}, 7981 (2014).
\bi{rev007} A. P. Gantapara, J. de Graaf, R. van Roij, and M. Dijkstra,
J. Chem. Phys. {\bf 142}, 05904 (2015).

\bi{rLR01} L. Radzihovsky, and T. C. Lubensky,
EuroPhys. Lett. {\bf 54}, 206 (2001).

\bi{rLR02} T. C. Lubensky, and L. Radzihovsky,
Phys. Rev. E {\bf 66}, 031704 (2002).

\bi{rBCP01}
P. E. Cladis, H. R. Brand, and H. Pleiner.
Liquid Crystals Today, {\bf 9}, 1 (1999).

\bi{rBCP02}
H. R. Brand, P. E. Cladis, and H. Pleiner,
EuroPhys. Lett. {\bf 57}, 368 (2002).

\bi{rBCP03}
H. R. Brand, P. E. Cladis, and H. Pleiner,
Ferroelectrics {\bf 315}, 165 (2005).

\bi{rtetsim01}
S. Romano, Phys. Rev. E {\bf 77}, 021704 (2008).

\bi{rtetsim02}
L. Longa, G. Paj\c{a}k, and T. Wydro,
Phys. Rev. E {\bf 79}, 040701(R) (2009).

\bi{rtetsim03}
K. Trojanowski, G. Pai\c{a}k, L. Longa, and T. Wydro,
Phys. Rev. E {\bf 86}, 011704 (2012).

\bi{rbanrev} G. Pelzl, S. Diele, and W. Weissflog,
Adv. Mater. {\bf 11}, 707 (1999).

\bi{rbannem} J. Matraszek, J. Mieczkowski, J. Szyd{\l}owska, and E. Gorecka,
Liq. Cryst. {\bf 27}, 429 (2000).

\bi{rtet03}
E. Wiant, K. Neupane, S. Sharma, J. T. Gleeson, S. Sprunt,
A. J\'akli, N. Pradhan, and G. Iannacchione,
Phys. Rev. E {\bf 77}, 061701 (2008)

\bi{rcu00} J. A. C. Veerman, and D. Frenkel,
Phys. Rev. A {\bf 45}, 5632 (1992).
\bi{rcu01} A. Chamoux, and A. Perera,
J. Chem. Phys. {\bf 108}, 8172 (1998).
\bi{rcu02} A. Chamoux, and A. Perera,
Phys. Rev. E {\bf 58}, 1933 (1998).
\bi{rcu03} R. Blaak, D. Frenkel, and B. M. Mulder,
J. Chem. Phys. {\bf 110}, 11652 (1999).
\bi{rcu04} R. Blaak, and B. M. Mulder,
Phys. Rev. E {\bf 58}, 5873 (1998).
\bi{rcu04add} R. Blaak, B. M. Mulder, and D. Frenkel,
J. Chem. Phys. {\bf 120}, 5486 (2004).
\bi{rcu05} B. S. John, A. Stroock, and F. A. Escobedo,
J. Chem. Phys. {\bf 120}, 9383 (2004).
\bi{rcu06} B. S. John, A. Stroock, and F. A. Escobedo,
J. Phys. Chem. B {\bf 109}, 23008 (2005).
\bi{rcu06add} B. S. John, C. Juhlin, and F. A. Escobedo,
J. Chem. Phys. {\bf 128}, 044909 (2008).
%
\bi{pre074011704}
S. Romano, Phys. Rev. E {\bf 74}, 011704 (2006).
%
\bi{rcu07}
P. D. Duncan, M. Dennison, A. J. Masters, and M. R. Wilson,
Phys. Rev. E {\bf 79}, 031702 (2009).
\bi{rcu08}
P. D. Duncan, A. J. Masters, and M. R. Wilson,
Phys. Rev. E {\bf 84}, 011702 (2011).
\bi{rcu09}
M. Marechal, A. Patti, M. Dennison, and M. Dijkstra,
Phys. Rev. Lett. {\bf 108}, 206101 (2012).

\bi{rcu10}
M. R. Wilson, P. D. Duncan, M. Dennison, and A. J. Masters,
Soft Matter {\bf 8}, 3348 (2012).



\bi{rLV} H. N. W. Lekkerkerker, and G. J. Vroege,
Phil. Trans. R. Soc. A {\bf 371}, 20120263 (2013).

\bi{rQKR} S. J. S. Qazi, G. Karlsson, and A. R. Rennie,
J. Colloid Interface Sci. {\bf 348}, 80 (2010).

\bibitem{rLL01}
P. A. Lebwohl and G. Lasher, 
Phys. Rev. A  {\bf 5} 1350 (1972).

\bibitem{rLL02}
G. Lasher, 
Phys. Rev. A {\bf 6} 426 (1972).

\bi{bpz-05}
P. Pasini, C. Chiccoli, and  C. Zannoni,

{\it Advances in the Computer Simulations of Liquid Crystals},

ed. by P. Pasini and C. Zannoni, NATO Science Series, vol. C 545,

Kluwer, (Dordrecht, 2000), ch. 5. 

\bi{rBL} M. A. Bates, and G. R. Luckhurst,
Phys. Rev. E {\bf 72}, 051702 (2005).
\bibitem{rms}
W. Maier and A. Saupe, 
Z. Naturforsch. A {\bf 13}, 564 (1958); 
{\bf 14}, 882 (1959);
{\bf 15} 287 (1960).

\bi{rmfb}
G. R. Luckhurst,
{\it The Molecular Physics of Liquid Crystals}, 
ed. by G. R. Luckhurst and G. W. Gray, 
(Academic Press, London 1979),   
chap. 4, p. 85-120.

\bi{bqz-00}
G. R. Luckhurst, 

{\it Physical Properties of Liquid Crystals: Nematics},

ed by D. A. Dunmur, A. Fukuda,  G. R. Luckhurst,

INSPEC, (London, UK, 2001), ch.  2.1.



\bi{rimfAA}
C. G. Gray and K. E. Gubbins,
{\it Theory of Molecular Fluids, volume 1: Fundamentals},
Oxford University Press, (Oxford, UK, 1984).
\bi{rimfBB}
A. J. Stone, {\it The Theory of Intermolecular Forces},
Oxford University Press, Oxford, UK, 1997.
\bi{rdispa}
F. London, 
Trans. Faraday Soc. {\bf 33}, 8 (1937).
\bi{rdispb}
J. H. de Boer and G. Heller,
Physica {\bf 4}, 1045 (1937).
\bi{rdispc}
J. de Boer,
Physica {\bf 9}, 363 (1942).
\bi{rdispd}
A. J. van der Merwe,
Z. Phys. {\bf 196}, 212; 332 (1966).
\bi{rdex00}
B. C. Kohin, J. Chem. Phys. {\bf 33}, 882 (1960).
\bi{rdex01}
S. L. Price and A. J. Stone, 
Mol. Phys. {\bf 40}, 805 (1980).
\bi{rdex03}
E. Burgos, C. S. Murthy, and R. Righini,
Mol. Phys. {\bf 47}, 1391 (1982).

\bi{robj}
R. A. Kromhout, B. Linder,
J. Phys. Chem. {\bf 99}, 16909 (1995).

\REM{ 
\bi{rimf00}
A. D. Buckingham, in
{\rm Intermolecular Forces}, 
Adv. Chem. Phys. {\bf 12}, (1967),
edited by J. O. Hirschfelder  (Chapter 2). 

} 





\bi{rdisp1}
R. L. Humphries, G. R. Luckhurst, and S. Romano,
Mol. Phys. {\bf 42}, 1205 (1981).

\bi{PSthesis} P. Simpson, Ph. D. thesis, Southampton University, UK (1982).

\bi{rdisp2}
S. Romano,
Liq. Cryst. {\bf 3}, 323 (1988).

\bi{rBates}
M. A. Bates, Phys. Rev. E {\bf 65}, 041706 (2002). 

\bi{r2ddisp} S. Romano, Physica A {\bf 322}, 432 (2003).

\bi{rsi00}
J. Nehring and A. Saupe,
J. Chem. Phys. {\bf 54}, 337 (1971); {\it ibid}. {\bf 56}, 5527 (1972).

\bi{rJeu}
G. Vertogen and W. H. de Jeu,
{\it Thermotropic Liquid Crystals, Fundamentals},
Springer Verlag, (Berlin, 1988).

\bi{rBBel} 
G. Barbero and R. Barberi,
 {\it Physics of Liquid Crystalline Materials},
ed. by I.-C. Khoo and F. Simoni, Gordon and Breach
(Philadelphia, 1991);

chapter 8, pp. 183-213.

\bi{rHR} R. Hashim, and S. Romano, Int. J. Mod. Phys. B {\bf 13}, 3879 (1999).

\bi{rselflgns}
S. Romano,
Mod. Phys. Lett. B {\bf 15}, 137 (2001).


\bi{r2dns}
S. Romano, 
Phys. Lett. A {\bf 305}, 196 (2002).
%

\bi{rADB} A. D. Buckingham, Disc. Faraday Soc. {\bf 43}, 205 (1967).

\bi{r23}
C. Zannoni,
{\it The Molecular Physics of Liquid Crystals}, 
ed. by G. R. Luckhurst and G. W. Gray, 
(Academic Press, London 1979), 
chap. 3, p. 51-84.
\bi{r24}
C. Zannoni, 
{\it The Molecular Physics of Liquid Crystals}, 
ed. by G. R. Luckhurst and G. W. Gray, (Academic Press, London 1979),  
chap. 9, p. 191-220.
\bi{bpz-02}
C. Zannoni,

{\it Advances in the Computer Simulations of Liquid Crystals},

ed. by P. Pasini and C. Zannoni, NATO Science Series, vol. C 545,

Kluwer, (Dordrecht, 2000), ch. 2. pp. 17-50. 

\bi{SST} S. S. Turzi, J. Math. Phys. {\bf 52}, 053517 (2011).

\bi{ref00}
G. B. Arfken, H. J. Weber,

{\it Mathematical Methods for Physicists}, 4th edition,

Academic Press, (San Diego, USA, 1995).


\bi{rLN}
J. W. Leech, and D. J. Newman,
{\it How to Use Groups}, Methuen, (London, 1969).  

\bi{rrr001}
K. Zheng and P. Palffy-Muhoray,
Electronic-Liquid Crystal Communications (2007), 
\texttt{http://www.e-lc.org/docs/2007\_02\_03\_02\_33\_15}.

\bi{rrr002} J. Geng, and J. V. Selinger, Phys. Rev. E {\bf 80}, 011707 (2009).
%
\bi{rrr003} P. A. Lebwohl, and G. Lasher, Phys. Rev. A {\bf 6}, 426 (1972).
%
\bi{rrr004} U. Fabbri, and C. Zannoni, Mol. Phys. {\bf 58}, 763 (1986).
%
\bi{rrr005} C. Chiccoli, P. Pasini, and C. Zannoni, Physica A {\bf 148}, 298 (1988).
%
\bi{rrr006} J. T. Chalker, P. C. W. Holdsworth, and E. F. Shender,
Phys. Rev. Lett. {\bf 68}, 855 (1992).
\bi{rrr007} R. Moessner, and J. T. Chalker,
Phys. Rev. B {\bf 58}, 12049 (1998).
\bi{rrr008} J. M. Hopkinson, S. V. Isakov, H.-Y. Kee, and Y. B. Kim,
Phys. Rev. Lett. {\bf 99}, 037201 (2007).
%
%
\bi{rq01} J. Felsteiner, D. B. Litvin, and J. Zak,
Phys. Rev. B {\bf 3}, 2706 (1971).
\bi{rq02} J. C. Raich, J. Chem. Phys. {\bf 56}, 2395 (1972).
\bi{rq03} J. C. Raich, and R. D. Etters, J. Low Temp, Phys. {\bf 7}, 449 (1972).
\bi{rq04} J. Felsteiner, and Z, Friedman, 
Phys. Rev. B {\bf 8}, 3996 (1973). 
\bi{rq05} M. J. Mandell, J. Chem. Phys. {\bf 60}, 1432 (1974); ibid. {\bf 60}, 	4880 (1974).
\bi{rq06} S. Romano, Z. Naturforsch. A {\bf 29}, 1631 (1974).
\bi{rq07} T. A. Scott, Phys. Rep. {\bf 27}, 85 (1976).
\bi{rq08} see, for example, \texttt{http://rruff.geo.arizona.edu/AMS/minerals/Nitrogen}.
\end{thebibliography}
\end{document}